\documentclass[10pt,a4paper]{article}
\usepackage{amsfonts,amsmath,amssymb,cite,lmodern}
\usepackage[mathscr]{euscript}
\usepackage[colorlinks=true,citecolor=red,linkcolor=red,%
urlcolor=blue]{hyperref}
\usepackage{graphicx}
\usepackage{subcaption}
\usepackage{float}
\newcommand{\sgn}{\mathop{\textrm{sgn}\mskip1mu}\limits}
\newcommand{\Tr}{\mathop{\textrm{Tr}}}

\begin{document}
\title{Dynamical quasi-symmetry of the classical Kepler--Coulomb
problem and the conservation of the Laplace--Runge--Lenz vector}
\author{Rados{\l}aw Szmytkowski\footnote{Email:
radoslaw.szmytkowski@pg.edu.pl} \\*[3ex]
Faculty of Applied Physics and Mathematics, \\
Gda{\'n}sk University of Technology, \\
ul.\ Gabriela Narutowicza 11/12, 80--233 Gda{\'n}sk, Poland}
%
%
\maketitle
\begin{abstract}
The paper studies a classical non-relativistic particle moving in
$\mathbb{R}^{N}$, \mbox{$(N\geqslant2)$}, in the presence of the
Kepler--Coulomb force
\mbox{$\boldsymbol{F}(\boldsymbol{r})=\alpha\boldsymbol{r}/r^{3}$},
\mbox{$(\alpha\neq0)$}. It is shown that the
extended-configuration-space transformation involving the projective
transformation of the radius vector
\mbox{$\boldsymbol{r}\to\boldsymbol{R}
=\boldsymbol{r}/(1+\boldsymbol{c}\cdot\boldsymbol{r})$} and the
differential time scaling \mbox{$\mathrm{d}t\to\mathrm{d}T
=\mathrm{d}t/(1+\boldsymbol{c}\cdot\boldsymbol{r})^{2}$}, where
$\boldsymbol{c}$ is a real vector-valued parameter with dimensions of
inverse length, constitutes a dynamical quasi-symmetry in the sense
that it maps Newton's equation of motion
\mbox{$m(\mathrm{d}^{2}\boldsymbol{r}/\mathrm{d}t^{2})
=\alpha\boldsymbol{r}/r^{3}$} into the equation
\mbox{$m(\mathrm{d}^{2}\boldsymbol{R}/\mathrm{d}T^{2})
=(\alpha\boldsymbol{R}/R^{3})
\sgn(1-\boldsymbol{c}\cdot\boldsymbol{R})$}. This quasi-symmetry is
independent of, yet complementary to, the well-known Kepler--Coulomb
dynamical symmetry, which is SO($N+1$) for negative energies and
SO($N,1$) for positive energies. A detailed analytical and graphical
analysis of induced mappings between non-degenerate conic
trajectories associated with the original and transformed systems is
presented. Mappings between velocity hodographs are also briefly
discussed. It is then proved that the conservation of energy in the
transformed motion implies the conservation of the
Laplace--Runge--Lenz vector in the original dynamics.
\end{abstract}
%
%
%
\section{Introduction}
\label{I}
\setcounter{equation}{0}
Studies on dynamical symmetries of the classical non-relativistic
Kepler--Coulomb problems in three, or even more generally
in $N$, spatial dimensions have a relatively long history and are
extensively documented in the literature. A survey of relevant works
indicates that, to date, most research focused on issues related to
the SO$(N+1)$ and SO$(N,1)$ symmetry groups associated with negative
and positive values of the total energy of a particle, respectively,
as well as on certain generalizations thereof (see, e.g., Gy{\"o}rgyi
1968, 1969; Belbruno 1977; Osipov 1977; Milnor 1983; Cordani 2003;
Wulfman 2011, and references therein). Among several exceptions known
to the present author that address other possible types of symmetries
are a section of a review by Mariwalla (1975, Sec.\/~5.3), papers by
Cari{\~n}ena et al.\ (1991), Krause (1994), and Bor and Jackman
(2023), as well as a section of a textbook by Kholshevnikov and Titov
(2007, Sec.\/~2.5). Of these, Mariwalla and Krause independently
claimed that Newton's equation of motion for a non-relativistic
particle moving in the field of the Kepler--Coulomb inverse-square
force is invariant under the composite transformation
$(\boldsymbol{r},t)\to(\boldsymbol{R},T)$ consisting of the
projective mapping of the particle's radius vector
\begin{subequations}
\begin{equation}
\boldsymbol{r}\to\boldsymbol{R}=\frac{\boldsymbol{r}}
{1+\boldsymbol{c}\cdot\boldsymbol{r}}
\label{1.1a}
\end{equation}
and the differential time scaling\footnote{~We have traced the use of
the differential time scaling
\mbox{$\mathrm{d}T=\rho(\boldsymbol{r})\mathrm{d}t$} (also referred
to in the literature as the infinitesimal time reparametrization or,
perhaps somewhat undeservedly, the generalized Sundman
transformation) in mechanics back to several French and Italian
papers from the end of the 19th century. The earliest of these are
two papers by Appell (1889a, 1889b). Perhaps the best known among
such time reparametrizations is the Levi-Civita--Sundman
transformation $\mathrm{d}T=(\beta/r)\mathrm{d}t$,
($\beta\in\mathbb{R}_{+}$), introduced at the beginning of the 20th
century and widely employed ever since in celestial mechanics for the
regularization of the gravitational three-body problem.}
\begin{equation}
\mathrm{d}t\to\mathrm{d}T=\frac{\mathrm{d}t}
{(1+\boldsymbol{c}\cdot\boldsymbol{r})^{2}},
\label{1.1b}
\end{equation}
\label{1.1}%
\end{subequations}
where $\boldsymbol{c}$ is a real position- and time-independent
vector with dimensions of inverse length that parametrizes the
transformations. Particularly elegant was the approach adopted by
Krause, who arrived at the transformations in Eqs.\ (\ref{1.1a}) and
(\ref{1.1b}) by generalizing Lie's method for constructing continuous
symmetries of differential equations.

The present study has two goals. The first is to point out that,
contrary to the claims of Mariwalla and Krause, the transformations
in Eqs.\ (\ref{1.1a}) and (\ref{1.1b}) do \emph{not\/} constitute a
symmetry of Newton's equation for the Kepler--Coulomb problem. We
shall show that this is because the original and transformed
dynamical equations, once the latter is correctly derived (which both
Mariwalla and Krause failed to do), turn out \emph{not\/} to be
globally identical in form. Nevertheless, we shall argue that the
transformations under consideration may be regarded as a kind of
\emph{quasi}-symmetry of the corresponding Newton's equation. The
second goal is to study how this quasi-symmetry manifests itself in
the dynamical characteristics of the Kepler--Coulomb problem and to
analyze its consequences for the relevant conservation laws.

The remainder of the paper is organized as follows. In
Sec.\/~\ref{II}, we investigate in detail the effect of the
transformations (\ref{1.1a}) and (\ref{1.1b}) on Newton's equations
of motion for a particle driven by the Kepler--Coulomb force. Section
\ref{III} contains an in-depth analytical and partly graphical
analysis of how the Kepler--Coulomb orbits and velocity hodographs
map under the transformations considered. The similarities and
differences between our approach and that of Kholshevnikov and Titov
(2007, Sec.\/~2.5) are discussed there as well. In Sec.\/~\ref{IV},
we prove that, with a suitable restriction imposed on the domain of
the vector parameter $\boldsymbol{c}$, the well-known conservation
law for the Laplace--Runge--Lenz vector in the Kepler--Coulomb field
may be derived from the energy conservation law for the transformed
motion. Concluding remarks are presented in Sec.\/~\ref{V}.
%
%
\section{Quasi-symmetry of Newton's equation of motion for the
Kepler--Coulomb problem}
\label{II}
\setcounter{equation}{0}
Consider the Kepler--Coulomb problem, in which a classical point
particle of constant mass $m$ moves in an $(N\geqslant2)$-dimensional
Euclidean space under the action of the inverse-square central force
\begin{equation}
\boldsymbol{F}(\boldsymbol{r})=\alpha\frac{\boldsymbol{r}}{r^{3}}
\qquad (\alpha\neq0).
\label{2.1}
\end{equation}
Here $\boldsymbol{r}$ is the position vector of the particle relative
to the force center, while $r=|\boldsymbol{r}|$. Newton's equation
governing the particle's dynamics is
\begin{equation}
m\frac{\mathrm{d}\boldsymbol{v}}{\mathrm{d}t}
=\alpha\frac{\boldsymbol{r}}{r^{3}},
\label{2.2}
\end{equation}
where $\boldsymbol{v}=\mathrm{d}\boldsymbol{r}/\mathrm{d}t$ is the
particle's velocity. The angular momentum tensor $\hat{\mathbf{L}}$
is defined in terms of dyadic products of the vectors
$\boldsymbol{r}$ and $\boldsymbol{v}$ as
\begin{equation}
\hat{\mathbf{L}}=m\left(\boldsymbol{r}\boldsymbol{v}
-\boldsymbol{v}\boldsymbol{r}\right).
\label{2.3}
\end{equation}
From Eqs.\ (\ref{2.2}) and (\ref{2.3}), it is easy to deduce that
\begin{equation}
\frac{\mathrm{d}\hat{\mathbf{L}}}{\mathrm{d}t}=m\left(\boldsymbol{r}
\frac{\mathrm{d}\boldsymbol{v}}{\mathrm{d}t}
-\frac{\mathrm{d}\boldsymbol{v}}{\mathrm{d}t}\boldsymbol{r}\right)
=\hat{\mathbf{0}},
\label{2.4}
\end{equation}
where $\hat{\mathbf{0}}$ stands for the zero tensor. This means that
the tensor $\hat{\mathbf{L}}$ is the constant of motion.

Consider now a projective transformation that maps the point with
the position vector $\boldsymbol{r}$ to the point with the position
vector $\boldsymbol{R}$ according to Eq.\ (\ref{1.1a}). The inverse
transformation is clearly given by
\begin{equation}
\boldsymbol{r}=\frac{\boldsymbol{R}}
{1-\boldsymbol{c}\cdot\boldsymbol{R}}.
\label{2.5}
\end{equation}
For later use, from Eq.\ (\ref{2.5}) we deduce the identity
\begin{equation}
(1+\boldsymbol{c}\cdot\boldsymbol{r})
(1-\boldsymbol{c}\cdot\boldsymbol{R})=1,
\label{2.6}
\end{equation}
from which it follows that
\begin{equation}
\sgn(1+\boldsymbol{c}\cdot\boldsymbol{r})
=\sgn(1-\boldsymbol{c}\cdot\boldsymbol{R})
\label{2.7}
\end{equation}
[throughout this paper, we adopt the convention that
\mbox{$\sgn(0^{\pm})=\pm1$}]. Differentiating Eq.\ (\ref{2.5}) with
respect to time $t$, we obtain
\begin{equation}
\boldsymbol{v}
=\frac{\displaystyle\frac{\mathrm{d}\boldsymbol{R}}{\mathrm{d}t}
+\boldsymbol{R}
\left(\boldsymbol{c}\cdot\frac{\mathrm{d}\boldsymbol{R}}
{\mathrm{d}t}\right)-\frac{\mathrm{d}\boldsymbol{R}}{\mathrm{d}t}
(\boldsymbol{c}\cdot\boldsymbol{R})}
{(1-\boldsymbol{c}\cdot\boldsymbol{R})^{2}}.
\label{2.8}
\end{equation}
At this point, it is convenient to recall the pseudo-time variable
$T$, defined earlier in terms of the physical time $t$ through the
differential relation (\ref{1.1b}). The inverse of Eq.\ (\ref{1.1b})
is
\begin{equation}
\mathrm{d}t=\frac{\mathrm{d}T}
{(1-\boldsymbol{c}\cdot\boldsymbol{R})^{2}}.
\label{2.9}
\end{equation}
By plugging Eq.\ (\ref{2.9}) into Eq.\ (\ref{2.8}) and defining
the pseudo-velocity
\begin{equation}
\boldsymbol{V}\equiv\frac{\mathrm{d}\boldsymbol{R}}{\mathrm{d}T},
\label{2.10}
\end{equation}
we cast Eq.\ (\ref{2.8}) into the form
\begin{equation}
\boldsymbol{v}=\boldsymbol{V}
+\boldsymbol{R}(\boldsymbol{c}\cdot\boldsymbol{V})
-\boldsymbol{V}(\boldsymbol{c}\cdot\boldsymbol{R}).
\label{2.11}
\end{equation}
From this, with the aid of Eq.\ (\ref{2.5}), we infer that
\begin{equation}
\boldsymbol{r}\boldsymbol{v}-\boldsymbol{v}\boldsymbol{r}
=\boldsymbol{R}{\mskip1mu}\boldsymbol{V}
-\boldsymbol{V}\!\boldsymbol{R},
\label{2.12}
\end{equation}
which, after being combined with Eq.\ (\ref{2.3}), yields
\begin{equation}
m(\boldsymbol{R}{\mskip1mu}\boldsymbol{V}
-\boldsymbol{V}\!\boldsymbol{R})=\hat{\mathbf{L}}.
\label{2.13}
\end{equation}
This shows that the angular momentum tensor $\hat{\mathbf{L}}$ is an
invariant of the transformation considered here. Equation
(\ref{2.11}) may be rewritten as
\begin{equation}
\boldsymbol{v}=\boldsymbol{V}+(\boldsymbol{R}{\mskip1mu}\boldsymbol{V}
-\boldsymbol{V}\!\boldsymbol{R})\cdot\boldsymbol{c},
\label{2.14}
\end{equation}
and this, by employing Eq.\ (\ref{2.13}), may be further simplified
to
\begin{equation}
\boldsymbol{v}=\boldsymbol{V}
+\frac{\hat{\mathbf{L}}\cdot\boldsymbol{c}}{m}.
\label{2.15}
\end{equation}
Differentiating Eq.\ (\ref{2.15}) with respect to time $t$, and then
using Eqs.\ (\ref{2.4}) and (\ref{2.9}), we find that
\begin{equation}
\frac{\mathrm{d}\boldsymbol{v}}{\mathrm{d}t}
=(1-\boldsymbol{c}\cdot\boldsymbol{R})^{2}
\frac{\mathrm{d}\boldsymbol{V}}{\mathrm{d}T}.
\label{2.16}
\end{equation}
When this relationship is inserted into Eq.\ (\ref{2.2}), and if
simultaneously Eq.\ (\ref{2.5}) is used to express the vector
$\boldsymbol{r}/r^{3}$ in terms of the vector $\boldsymbol{R}$ and
its modulus $R=|\boldsymbol{R}|$, we arrive at Newton's equation of
motion in the variables $(\boldsymbol{R},\boldsymbol{V},T)$:
\begin{equation}
m\frac{\mathrm{d}\boldsymbol{V}}{\mathrm{d}T}
=\alpha\frac{\boldsymbol{R}}{R^{3}}
\sgn(1-\boldsymbol{c}\cdot\boldsymbol{R}).
\label{2.17}
\end{equation}

On comparing Eqs.\ (\ref{2.2}) and (\ref{2.17}), we see that the
presence of the factor
\mbox{$\sgn(1-\boldsymbol{c}\cdot\boldsymbol{R})$} on the right-hand
side of the latter prevents the transformations (\ref{1.1a}) and
(\ref{1.1b}) from being a symmetry of Newton's equation of motion for
the Kepler--Coulomb problem. The contrary claim of Mariwalla (1975,
Sec.\/~5.3) arose from overlooking the signum factor; this, in turn,
resulted from the erroneous assumption that, within the notation
adopted here, the force (\ref{2.1}) satisfies
\begin{equation}
\boldsymbol{F}(\boldsymbol{R})
=(1+\boldsymbol{c}\cdot\boldsymbol{r})^{2}
\boldsymbol{F}(\boldsymbol{r})
\qquad (\textrm{incorrect!}),
\label{2.18}
\end{equation}
whereas in fact
\begin{equation}
\boldsymbol{F}(\boldsymbol{R})
=\sgn(1+\boldsymbol{c}\cdot\boldsymbol{r})
(1+\boldsymbol{c}\cdot\boldsymbol{r})^{2}
\boldsymbol{F}(\boldsymbol{r}).
\label{2.19}
\end{equation}
Krause (1994) arrived at essentially the same conclusion as
Mariwalla, although the methods used by the two authors to derive the
incorrect form of the transformed Newton's equation, with the signum
factor omitted, differed in detail. The critique of Krause's
reasoning is given elsewhere (Szmytkowski 2026).\footnote{~We find it 
striking that exactly the same type of error that we identified in 
the works of Mariwalla and Krause had been made several decades 
earlier by Appell in his investigation of the transformation of 
Newton's equations of motion in the Cartesian representation under 
general homographies in $\mathbb{R}^{2}$:\ the factor 
$\sgn(Cx_{1}+C'y_{1}+C'')$ is missing from the right-hand side of the 
bottom equation on page 107 of Appell (1889b).}

Let us examine Eq.\ (\ref{2.17}) more closely. As long as the
particle's original motion remains confined to the half-space
\mbox{$1+\boldsymbol{c}\cdot\boldsymbol{r}>0$} (containing the force
center at \mbox{$\boldsymbol{r}=\boldsymbol{0}$)}, the transformed
motion, according to Eq.\ (\ref{2.7}), takes place in the half-space
\mbox{$1-\boldsymbol{c}\cdot\boldsymbol{R}>0$} (containing the
transformed force center at
\mbox{$\boldsymbol{R}=\boldsymbol{0}$}); it then follows from Eqs.\
(\ref{2.2}) and (\ref{2.17}) that both the original and transformed
motions are governed by \emph{the same\/} Newton's equation, so that
in this particular case the dynamical symmetry is present. If,
however, during a given time interval the original motion takes place
in the half-space \mbox{$1+\boldsymbol{c}\cdot\boldsymbol{r}<0$},
then, again by Eq.\ (\ref{2.7}), the transformed motion is realized
in the half-space \mbox{$1-\boldsymbol{c}\cdot\boldsymbol{R}<0$}, and
the two motions are driven by forces of the same magnitude but
\emph{opposite\/} spatial orientation; in this case, we have a kind
of dynamical antisymmetry. In view of this, it seems justified to
conclude that, taken as a whole, the transformations (\ref{1.1a}) and
(\ref{1.1b}) constitute a dynamical \emph{quasi}-symmetry of Newton's
equation of motion for the Kepler--Coulomb problem.

Before proceeding to the next section, we briefly comment on the
group-theoretical classification of the transformations (\ref{1.1a})
and (\ref{1.1b}). Taken separately, the family of transformations of
type (\ref{1.1a}) forms an $N$-parameter unipotent Abelian subgroup
of the general projective group PGL$(N+1,\mathbb{R})$; this subgroup
is isomorphic to the translation group Trans$(N,\mathbb{R})$. Taken
together, the transformations of types (\ref{1.1a}) and (\ref{1.1b})
also form an $N$-parameter Abelian group, which again is isomorphic
to Trans$(N,\mathbb{R})$.
%
%
\section{Mapping of the Kepler--Coulomb orbits and velocity
hodographs}
\label{III}
\setcounter{equation}{0}
\subsection{Preliminaries on the Kepler--Coulomb orbits and velocity
hodographs}
\label{III.1}
This section reviews the key facts concerning orbits and velocity
hodographs in the Kepler--Coulomb problem and sets up the notation
for subsequent use.

Since the angular momentum tensor $\hat{\mathbf{L}}$ is conserved in
the force field (\ref{2.1}), the particle's motion is planar for any
spatial dimension $N\geqslant2$ and the plane of motion contains the
force center (Wheeler 2005, {\"O}nder 2006). Introducing the polar
coordinates $r$ and $\varphi$ in this plane, with the pole located at
the force center, and integrating Newton's equation (\ref{2.2}),
after suitable angular alignment one obtains the following polar 
representation of admissible orbits:
\begin{equation}
r=\frac{p}{e\cos\varphi-\sgn\alpha}.
\label{3.1}
\end{equation}
Here
\begin{equation}
p=\frac{L^{2}}{m|\alpha|},
\qquad
e=\sqrt{1+\frac{2EL^{2}}{m\alpha^{2}}}=\sqrt{1+\frac{2Ep}{|\alpha|}},
\label{3.2}
\end{equation}
where
\begin{equation}
E=\frac{1}{2}m\boldsymbol{v}^{2}+\frac{\alpha}{r}
\label{3.3}
\end{equation}
is the total particle energy and
\begin{equation}
L^{2}=\frac{1}{2}
\Tr\big(\hat{\mathbf{L}}\mbox{}^{\mathrm{T}}\cdot\hat{\mathbf{L}}\big)
=\frac{1}{2}
\Tr\big(\hat{\mathbf{L}}\cdot\hat{\mathbf{L}}\mbox{}^{\mathrm{T}}\big)
=-\frac{1}{2}\Tr\big(\hat{\mathbf{L}}\cdot\hat{\mathbf{L}}\big)
=m^{2}\big[\boldsymbol{r}^{2}\boldsymbol{v}^{2}
-(\boldsymbol{r}\cdot\boldsymbol{v})^{2}\big]
\label{3.4}
\end{equation}
(the superscript $\mathrm{T}$ denotes the transpose) represents the
squared norm of the particle's angular momentum tensor
$\hat{\mathbf{L}}$. Depending on $\alpha$ and $e$, Eq.\
(\ref{3.1}) describes a particular conic section with a focus at the
point $\boldsymbol{r}=\boldsymbol{0}$ (which, as we recall, is the 
force center); specifically, the orbit is:
\begin{list}{}{}
\item[(a)] an ellipse for $\alpha<0$ and $0\leqslant e<1$; its
other
focus is located at the point where $r=2ep/(1-e^{2})$ and
$\varphi=\pi$,
\item[(b)] a parabola for $\alpha<0$ and $e=1$; its vertex is
located
at the point where $r=p/2$ and $\varphi=0$,
\item[(c)] a branch of a hyperbola for $e>1$; if $\alpha<0$, it is
the one that encloses the focus at $\boldsymbol{r}=\boldsymbol{0}$ 
(hereafter, we shall refer to this branch as the ``attractive'' one), 
whereas if $\alpha>0$, it is the one that encloses the other focus, 
which is located at the point where $r=2ep/(e^{2}-1)$ and $\varphi=0$ 
(hereafter, we shall refer to this branch as the ``repulsive'' one).
\end{list}
The parameters $e$ and $p$ are the eccentricity and the semi-latus
rectum of the relevant conic, respectively. The ranges allowed for
the angular variable $\varphi$ in Eq.\ (\ref{3.1}) follow from the
requirement that the denominator of the fraction on the right-hand
side must be non-negative:
\begin{equation}
e\cos\varphi-\sgn\alpha\geqslant0.
\label{3.5}
\end{equation}
By restricting $\varphi$ to the fundamental interval $[-\pi,\pi]$
(taken as closed at both ends to ensure that for $e=1$ both arms of
the parabola are treated symmetrically), one obtains
%
\begin{equation}
\left\{
\begin{array}{cll}
-\pi<\varphi\leqslant\pi &
\textrm{for} & \textrm{$\alpha<0$, \: $0\leqslant e<1$}, \\*[1ex]
-\pi\leqslant\varphi\leqslant\pi &
\textrm{for} & \textrm{$\alpha<0$, \: $e=1$}, \\*[1ex]
-\arccos\,(-1/e)\leqslant\varphi\leqslant\arccos\,(-1/e) &
\textrm{for} & \textrm{$\alpha<0$, \: $e>1$}, \\*[1ex]
-\arccos\,(1/e)\leqslant\varphi\leqslant\arccos\,(1/e) &
\textrm{for} & \textrm{$\alpha>0$, \: $e>1$}.
\end{array}
\right.
\label{3.6}
\end{equation}
%

One can introduce the Cartesian coordinates $x$ and $y$ in the plane
of motion using the standard relations
\begin{equation}
x=r\cos\varphi,
\qquad
y=r\sin\varphi.
\label{3.7}
\end{equation}
By combining Eq.\ (\ref{3.7}) with Eq.\ (\ref{3.1}), we obtain the
parametric Cartesian representation of the Kepler--Coulomb orbits in
the form
\begin{equation}
x=\frac{p\cos\varphi}{e\cos\varphi-\sgn\alpha},
\qquad
y=\frac{p\sin\varphi}{e\cos\varphi-\sgn\alpha},
\label{3.8}
\end{equation}
with the polar angle $\varphi$, varying according to Eq.\
(\ref{3.6}), as the parameter.

Eliminating $\varphi$ from Eq.\ (\ref{3.8}) yields the implicit
equation
\begin{equation}
(1-e^{2})x^{2}+y^{2}+2epx-p^{2}=0,
\label{3.9}
\end{equation}
which can be further cast to the canonical form, which is either
\begin{equation}
x+\frac{y^{2}}{2p}=\frac{p}{2}
\qquad (e=1)
\label{3.10}
\end{equation}
or
\begin{equation}
\frac{\displaystyle\left(x+\frac{ep}{1-e^{2}}\right)^{2}}
{\displaystyle\frac{p^{2}}{(1-e^{2})^{2}}}
+\frac{y^{2}}{\displaystyle\frac{p^{2}}{1-e^{2}}}=1
\qquad (e\neq1),
\label{3.11}
\end{equation}
depending on the value taken by the eccentricity $e$. It should be
noted that while Eq.\ (\ref{3.8}) together with Eq.\ (\ref{3.6})
represent the particle's orbit in the strict sense, Eq.\ (\ref{3.9})
does so only for $0\leqslant e\leqslant1$. For $e>1$, Eq.\
(\ref{3.9}) [and thus also Eq.\ (\ref{3.11})] represents the entire
hyperbola, which serves here as the underlying curve of the orbit;
the orbit itself is, as already noted, confined to one of the two
branches of this curve, depending on the sign of the force constant
$\alpha$. In this case, uniquely determining the orbit requires an
additional condition, which, for example, can be
\begin{equation}
\left\{
\begin{array}{ll}
\displaystyle x\leqslant\frac{p}{e+1} & (\alpha<0), \\*[2ex]
\displaystyle x\geqslant\frac{p}{e-1} & (\alpha>0).
\end{array}
\right.
\label{3.12}
\end{equation}
For later use, we also note that if $e>1$, then, regardless of the
sign of $\alpha$, the two asymptotes of the hyperbolic orbit are
\begin{equation}
y=\pm\sqrt{e^{2}-1}\,x\mp\frac{ep}{\sqrt{e^{2}-1}}.
\label{3.13}
\end{equation}

The parametric Cartesian representation of the particle's velocity
hodograph is obtained by differentiating Eq.\ (\ref{3.8}) with
respect to time $t$ and simplifying the result with the aid of Eqs.\
(\ref{3.1}) and (\ref{3.2}), and of the relation
\begin{equation}
L\equiv L_{xy}=m\left(x\frac{\mathrm{d}y}{\mathrm{d}t}
-y\frac{\mathrm{d}x}{\mathrm{d}t}\right)
=mr^{2}\frac{\mathrm{d}\varphi}{\mathrm{d}t}
\label{3.14}
\end{equation}
[it should be noted that $L$ can have either sign; the square of $L$
is defined by Eq.\ (\ref{3.4})]. This results in
\begin{equation}
v_{x}=\frac{\alpha}{L}\sin\varphi,
\qquad
v_{y}=\frac{e|\alpha|}{L}-\frac{\alpha}{L}\cos\varphi,
\label{3.15}
\end{equation}
where $\varphi$ ranges as in Eq.\ (\ref{3.6}). By eliminating
$\varphi$ from Eq.\ (\ref{3.15}), we obtain the identity
\begin{equation}
v_{x}^{2}+\left(v_{y}-\frac{e|\alpha|}{L}\right)^{2}
=\frac{\alpha^{2}}{L^{2}},
\label{3.16}
\end{equation}
which over the full range represents a circle centered at
$(v_{x},v_{y})=(0,e|\alpha|/L)$ with radius $|\alpha|/|L|$. For
$\alpha<0$ and $0\leqslant e\leqslant1$, the hodograph is the
entire circle (\ref{3.16}), whereas for $e>1$ it is an arc rather
than the full circle. It is worthwhile noting that, for $e>1$, the
two arcs corresponding to opposite signs of $\alpha$ but the
same values of $e$ and $L$ together form the entire circle.
%
%
\subsection{Mapping of the Kepler--Coulomb orbits}
\label{III.2}
\subsubsection{Description in Cartesian coordinates}
\label{III.2.1}
Let us introduce a Cartesian coordinate system in $\mathbb{R}^{N}$,
whose origin $O$ is located at the force center. The axes are
oriented such that the particle's plane of motion coincides with the
$Ox_{1}x_{2}$ plane and consequently the particle's position vector
$\boldsymbol{r}$ relative to $O$ has the Cartesian coordinates
$(x_{1},x_{2},0,\ldots,0)$. Hereafter, the coordinates $x_{1}$ and
$x_{2}$ will be identified with the coordinates $x$ and $y$,
respectively, defined in Sec.\ \ref{III.1}. Since Eq.\ (\ref{1.1a})
involves the vector parameter $\boldsymbol{c}=(c_{x},c_{y},\ldots)$
through the dot product $\boldsymbol{c}\cdot\boldsymbol{r}$ only, the
component of $\boldsymbol{c}$ perpendicular to the plane of motion
has no effect on the transformation in question. It then follows that
the position vector $\boldsymbol{R}$ of the image point has the
Cartesian coordinates $(X,Y,0,\ldots,0)$, where
\begin{equation}
X=\frac{x}{1+c_{x}x+c_{y}y},
\qquad
Y=\frac{y}{1+c_{x}x+c_{y}y}.
\label{3.17}
\end{equation}
By substituting the expressions from Eq.\ (\ref{3.8}) into Eq.\
(\ref{3.17}), one obtains the following Cartesian parametric
representation of the transformed curve:
\begin{subequations}
\begin{equation}
X=\frac{p\cos\varphi}
{(e+c_{x}p)\cos\varphi+c_{y}p\sin\varphi-\sgn\alpha},
\label{3.18a}
\end{equation}
\begin{equation}
Y=\frac{p\sin\varphi}
{(e+c_{x}p)\cos\varphi+c_{y}p\sin\varphi-\sgn\alpha}.
\label{3.18b}
\end{equation}
\label{3.18}%
\end{subequations}
Here, as in Eq.\ (\ref{3.8}), the parameter is the polar angle
$\varphi$ in the $Oxy$ plane; it varies in the range defined by Eq.\
(\ref{3.6}). Equations (\ref{3.18}) provide a convenient way of
generating plots of transformed curves. However, we defer the
presentation of such plots to Sec.\ \ref{III.2.3}, where each
admissible type of transformed curve will be characterized by the
number and nature of points shared by the original orbit and the
singular line of the projective transformation (\ref{3.17}).

Eliminating the parameter $\varphi$ from Eqs.\ (\ref{3.18}) yields
the implicit Cartesian equation
\begin{align}
& (1-e^{2}-2ec_{x}p-c_{x}^{2}p^{2})X^{2}+(1-c_{y}^{2}p^{2})Y^{2}
-2c_{y}p(e+c_{x}p)XY+2p(e+c_{x}p)X+2c_{y}p^{2}Y-p^{2}=0
\nonumber \\
&
\label{3.19}
\end{align}
of the underlying curve of the transformed curve. Alternatively, one
can arrive at Eq.\ (\ref{3.19}) by substituting into Eq.\ (\ref{3.9})
the expressions
\begin{equation}
x=\frac{X}{1-c_{x}X-c_{y}Y},
\qquad
y=\frac{Y}{1-c_{x}X-c_{y}Y}
\label{3.20}
\end{equation}
that are inverse to those in Eq.\ (\ref{3.17}). If $c_{x}=-e/p$ and
$c_{y}=0$, Eq.\ (\ref{3.19}) represents the circle centered at the
origin with radius $p$. In all other cases, to reduce Eq.\
(\ref{3.19}) to a more amenable form, we introduce the rotated
coordinates
\begin{equation}
X'=X\cos\Phi_{0}+Y\sin\Phi_{0},
\qquad
Y'=-X\sin\Phi_{0}+Y\cos\Phi_{0}.
\label{3.21}
\end{equation}
Choosing the angle $-\pi<\Phi_{0}\leqslant\pi$ such that
\begin{subequations}
\begin{equation}
\cos\Phi_{0}=\frac{e+c_{x}p}{\sqrt{(e+c_{x}p)^{2}+c_{y}^{2}p^{2}}}
\label{3.22a}
\end{equation}
and
\begin{equation}
\sin\Phi_{0}=\frac{c_{y}p}{\sqrt{(e+c_{x}p)^{2}+c_{y}^{2}p^{2}}},
\label{3.22b}
\end{equation}
\label{3.22}%
\end{subequations}
from Eq.\ (\ref{3.19}) we arrive at
\begin{equation}
(1-\varepsilon^{2})X^{\prime\mskip1mu2}+Y^{\prime\mskip1mu2}
+2\varepsilon pX'-p^{2}=0,
\label{3.23}
\end{equation}
where we have defined
\begin{equation}
\varepsilon=\sqrt{(e+c_{x}p)^{2}+c_{y}^{2}p^{2}}
\label{3.24}
\end{equation}
[for later applications, we observe that in the plane of the first
two coordinates of the vector $\boldsymbol{c}$, Eq.\ (\ref{3.24})
represents a circle centered at $(c_{x},c_{y})=(-e/p,0)$ with radius
$\varepsilon/p$]. Equation (\ref{3.23}) can be further cast into the
canonical form, which is either
\begin{equation}
X'+\frac{\displaystyle Y^{\prime\mskip1mu2}}{2p}=\frac{p}{2}
\qquad (\varepsilon=1)
\label{3.25}
\end{equation}
or
\begin{equation}
\frac{\displaystyle
\left(X'+\frac{\varepsilon p}{1-\varepsilon^{2}}\right)^{2}}
{\displaystyle\frac{p^{2}}{(1-\varepsilon^{2})^{2}}}
+\frac{\displaystyle Y^{\prime\mskip1mu2}}
{\displaystyle\frac{p^{2}}{1-\varepsilon^{2}}}=1
\qquad (\varepsilon\neq1),
\label{3.26}
\end{equation}
depending on the particular value taken by the parameter
$\varepsilon$.

It is now evident that the curve (\ref{3.19}) is a conic of
eccentricity $\varepsilon$ and of semi-latus rectum $p$; remarkably,
its semi-latus rectum coincides with that of the original orbit. The
conic has a focus located at $(X,Y)=(0,0)$, which is, as follows from
Eq.\ (\ref{2.17}), the center of the transformed force. The focal
axis of symmetry is rotated with respect to the $X$-axis by the angle
$\Phi_{0}$, defined by Eqs.\ (\ref{3.22}). Specifically:
\begin{enumerate}
\item If $0\leqslant\varepsilon<1$, i.e., if the point
$(c_{x},c_{y})$ lies in the open disk
\begin{equation}
\left(c_{x}+\frac{e}{p}\right)^{2}+c_{y}^{2}<\frac{1}{p^{2}},
\label{3.27}
\end{equation}
then the underlying conic (\ref{3.19}) is an ellipse with its other
focus at
\begin{equation}
(X,Y)=\left(-\frac{2p(e+c_{x}p)}{1-\varepsilon^{2}}, \,
-\frac{2c_{y}p^{2}}{1-\varepsilon^{2}}\right).
\label{3.28}
\end{equation}
\item If $\varepsilon=1$, i.e., if the point $(c_{x},c_{y})$ belongs
to the circle
\begin{equation}
\left(c_{x}+\frac{e}{p}\right)^{2}+c_{y}^{2}=\frac{1}{p^{2}},
\label{3.29}
\end{equation}
then the underlying conic (\ref{3.19}) is a parabola with its vertex
at
\begin{equation}
(X,Y)=\left(\frac{1}{2}p(e+c_{x}p),\,\frac{1}{2}c_{y}p^{2}\right).
\label{3.30}
\end{equation}
\item If $\varepsilon>1$, i.e., if the point $(c_{x},c_{y})$ lies in
the exterior
\begin{equation}
\left(c_{x}+\frac{e}{p}\right)^{2}+c_{y}^{2}>\frac{1}{p^{2}}
\label{3.31}
\end{equation}
of the closure of the disk (\ref{3.27}), then the underlying conic
(\ref{3.19}) is a hyperbola with its other focus at
\begin{equation}
(X,Y)=\left(\frac{2p(e+c_{x}p)}{\varepsilon^{2}-1},\,
\frac{2c_{y}p^{2}}{\varepsilon^{2}-1}\right)
\label{3.32}
\end{equation}
and with the asymptotes
%
\begin{subequations}
\begin{equation}
Y=\frac{\displaystyle c_{y}p(e+c_{x}p)\pm\sqrt{\varepsilon^{2}-1}}
{1-c_{y}^{2}p^{2}}X
\mp\frac{\displaystyle p\big[(e+c_{x}p)
\pm\sqrt{\varepsilon^{2}-1}\,c_{y}p\big]}
{\sqrt{\varepsilon^{2}-1}\,\big(1-c_{y}^{2}p^{2}\big)}
\qquad (|c_{y}|\neq1/p),
\label{3.33a}
\end{equation}
\begin{equation}
\left\{
\begin{array}{l}
\displaystyle
X=\sgn(e+c_{x}p)\frac{p}{\sqrt{\varepsilon^{2}-1}}, \\*[2ex]
\displaystyle
Y=-\sgn[c_{y}(e+c_{x}p)]
\frac{\varepsilon^{2}-2}{2\sqrt{\varepsilon^{2}-1}}X
+\sgn(c_{y})\frac{\varepsilon^{2}p}{2(\varepsilon^{2}-1)}
\end{array}
\right.
\qquad (|c_{y}|=1/p).
\label{3.33b}
\end{equation}
\label{3.33}%
\end{subequations}
%
\end{enumerate}

It turns out that the relation between the transformed curve
(\ref{3.18}) and its underlying conic (\ref{3.19}) is more involved
than in the case of the original orbit (\ref{3.8}) and its underlying
curve (\ref{3.9}). It is seen from Eqs.\ (\ref{3.18}) that, formally,
the Cartesian coordinates of the image point are $2\pi$-periodic
functions of the parameter $\varphi$. This implies that if $\varphi$
is allowed to vary in the interval being either $(-\pi,\pi]$ or
$[-\pi,\pi]$, which, as we know, corresponds to the case when the
original orbit is either an ellipse or a parabola, then the
transformed curve coincides with the entire underlying conic
(\ref{3.19}); in particular, if $e$, $c_{x}p$, and $c_{y}p$ are such
that $\varepsilon>1$, the transformed curve traces out \emph{both\/}
branches of a hyperbola. If, however, $\varphi$ varies over a range
that is a proper subset of the interval $(-\pi,\pi)$, which,
according to Eq.\ (\ref{3.6}), corresponds to the case when the
original orbit is one of the two branches of a hyperbola, then the
transformed curve forms only a part of the conic (\ref{3.19}):\
either a single arc of an ellipse, a parabola, or a hyperbola, or a
union of such arcs. This is clearly visible in the representative
plots of the transformed curves to be presented in Sec.\
\ref{III.2.3}.
%
%
\subsubsection{Description in polar coordinates}
\label{III.2.2}
Next, we turn to describing transformed curves in polar coordinates
in the $OXY$ plane. Since it holds that $R=\sqrt{X^{2}+Y^{2}}$, from
Eqs.\ (\ref{3.17}) and (\ref{3.7}) we find that
\begin{equation}
R=\frac{r}{|1+c_{x}r\cos\varphi+c_{y}r\sin\varphi|}.
\label{3.34}
\end{equation}
By defining the polar angle $\Phi$ via the relations
\begin{equation}
\cos\Phi=\frac{X}{R},
\qquad
\sin\Phi=\frac{Y}{R},
\label{3.35}
\end{equation}
and using Eqs.\ (\ref{3.17}), (\ref{3.7}), and (\ref{3.34}), we get
\begin{subequations}
\begin{equation}
\cos\Phi=\cos\varphi\sgn(1+c_{x}r\cos\varphi+c_{y}r\sin\varphi),
\label{3.36a}
\end{equation}
\begin{equation}
\sin\Phi=\sin\varphi\sgn(1+c_{x}r\cos\varphi+c_{y}r\sin\varphi).
\label{3.36b}
\end{equation}
\label{3.36}%
\end{subequations}
It follows from Eqs.\ (\ref{3.34}) and (\ref{3.36}) that whenever the
original orbit crosses the line
$1+c_{x}r\cos\varphi+c_{y}r\sin\varphi=0$ (i.e., the line
$1+c_{x}x+c_{y}y=0$), the radial coordinate $R$ of the image particle
tends to infinity, whereas the polar angle $\Phi$ jumps by $\pm\pi$.
Further, combining the relations (\ref{3.34}) and (\ref{3.36}) with
the identity
\begin{equation}
\sgn(1+c_{x}r\cos\varphi+c_{y}r\sin\varphi)
=\sgn(1-c_{x}R\cos\Phi-c_{y}R\sin\Phi),
\label{3.37}
\end{equation}
which follows from Eqs.\ (\ref{2.7}), (\ref{3.7}), and (\ref{3.35}),
and additionally using Eq.\ (\ref{3.1}), we obtain
%
\begin{equation}
R=\frac{p}{(e+c_{x}p)\cos\Phi+c_{y}p\sin\Phi
-\sgn\alpha\sgn(1-c_{x}R\cos\Phi-c_{y}R\sin\Phi)}.
\label{3.38}
\end{equation}
The above result can be written in a slightly more transparent form
by employing Eqs.\ (\ref{3.22}) and (\ref{3.24}), which yields
\begin{equation}
R=\frac{p}{\varepsilon\cos(\Phi-\Phi_{0})
-\sgn\alpha\sgn(1-c_{x}R\cos\Phi-c_{y}R\sin\Phi)}.
\label{3.39}
\end{equation}
Equation (\ref{3.39}) is manifestly implicit due to the factor
$\sgn(1-c_{x}R\cos\Phi-c_{y}R\sin\Phi)$ on the right-hand side, which
cannot be disentangled globally. However, this equation admits a
local disentanglement
\begin{equation}
R=\frac{p}{\varepsilon\cos(\Phi-\Phi_{0})\mp\sgn\alpha}
\qquad [\sgn(1-c_{x}R\cos\Phi-c_{y}R\sin\Phi)\gtrless0]
\label{3.40}
\end{equation}
%
in each of the two open half-planes into which the line
$1-c_{x}R\cos\Phi-c_{y}R\sin\Phi=0$ (i.e., the line
$1-c_{x}X-c_{y}Y=0$) divides the plane of transformed motion. For
both cases, Eq.\ (\ref{3.40}) describes a subset of a conic section
with eccentricity $\varepsilon$, semi-latus rectum $p$, and the focal
axis rotated by the angle $\Phi_{0}$ with respect to the polar ray
(coinciding with the positive semi-axis $OX$). This agrees with
results of our previous Cartesian-coordinate analysis.
%
%
\subsubsection{Classification of transformed curves by the number and
nature of the intersection points of the original orbit and the
singular line of the mapping (\ref{3.17})}
\label{III.2.3}
The purpose of this section is to show that the type of the underlying
conic (\ref{3.19}) of the transformed curve is directly related to
the number and nature of the points at which the singular line of the
projective transformation (\ref{3.17}) intersects the underlying
conic (\ref{3.9}) of the original orbit. The Cartesian coordinates of
these intersection points are solutions of the algebraic system
\begin{equation}
\left\{
\begin{array}{l}
(1-e^{2})x^{2}+y^{2}+2epx-p^{2}=0 \\*[1ex]
1+c_{x}x+c_{y}y=0.
\end{array}
\right.
\label{3.41}
\end{equation}
Below we analyze these solutions separately for the cases
$c_{y}\neq0$ and $c_{y}=0$.

Assume first that $c_{y}\neq0$. Solving the second of Eqs.\
(\ref{3.41}) for $y$ and inserting the result into the first one, we
obtain
\begin{equation}
\left(1-e^{2}+\frac{c_{x}^{2}}{c_{y}^{2}}\right)x^{2}
+2\left(ep+\frac{c_{x}}{c_{y}^{2}}\right)x
+\frac{1}{c_{y}^{2}}-p^{2}=0.
\label{3.42}
\end{equation}
The expression on the left-hand side of Eq.\ (\ref{3.42}) is
quadratic if the coefficient of $x^{2}$ is non-zero, linear if this
coefficient vanishes but the coefficient of $x$ is non-zero, and
constant if the coefficients of $x^{2}$ and $x$ both vanish. We shall
consider each of these cases in turn.

Let $1-e^{2}+c_{x}^{2}/c_{y}^{2}\neq0$. The discriminant of the
quadratic on the left-hand side of Eq.\ (\ref{3.42}) is
\begin{equation}
\Delta=4\,\frac{\varepsilon^{2}-1}{c_{y}^{2}},
\label{3.43}
\end{equation}
with $\varepsilon$ defined in Eq.\ (\ref{3.24}). Thus, we see that
the conic section (\ref{3.9}) and the line $1+c_{x}x+c_{y}y=0$ have
two distinct intersection points for $\Delta>0$ (which corresponds to
$\varepsilon>1$), a single point of tangency for $\Delta=0$ (which
corresponds to $\varepsilon=1$), or no points in common for
$\Delta<0$ (which corresponds to $0\leqslant\varepsilon<1$).

Let $1-e^{2}+c_{x}^{2}/c_{y}^{2}=0$ and $ep+c_{x}/c_{y}^{2}\neq0$.
The first of these conditions implies $e\geqslant1$, i.e., the conic
(\ref{3.9}) has to be either a parabola or a hyperbola, with
$c_{x}=0$ in the parabolic case $e=1$. Equation (\ref{3.42}) is then
linear in $x$, and its root corresponds to a single intersection
point of the line $1+c_{x}x+c_{y}y=0$ with the conic (\ref{3.9}). We
observe that in this case the difference $\varepsilon^{2}-1$ can be
written as
\begin{equation}
\varepsilon^{2}-1
=\left\{
\begin{array}{lcl}
c_{y}^{2}p^{2} && (e=1) \\*[1ex]
\displaystyle
\frac{(e^{2}-1+ec_{x}p)^{2}}{e^{2}-1} && (e>1).
\end{array}
\right.
\label{3.44}
\end{equation}
In both cases the resulting expressions are positive, implying that
$\varepsilon>1$.

Let $1-e^{2}+c_{x}^{2}/c_{y}^{2}=0$ and $ep+c_{x}/c_{y}^{2}=0$ hold
at the same time (clearly, this can occur only when $e>1$). Treating
these two conditions as an algebraic system in the unknowns $c_{x}$
and $c_{y}$, by solving it we find that
\begin{equation}
c_{x}=-\frac{e^{2}-1}{ep},
\qquad
c_{y}=\pm\frac{\sqrt{e^{2}-1}}{ep}.
\label{3.45}
\end{equation}
Referring to Eq.\ (\ref{3.13}), we see that this corresponds to the
situation in which the singular line \mbox{$1+c_{x}x+c_{y}y=0$}
coincides with either asymptote of the hyperbola, which is the
original conic (\ref{3.9}) for $e>1$. Inserting $c_{y}$ from Eq.\
(\ref{3.45}) into the constant term on the right-hand side of Eq.\
(\ref{3.42}), we obtain a contradiction, which is consistent with the
fact that the asymptotes of a hyperbola have no finite points in
common with it. Finally, we note that from Eqs.\ (\ref{3.24}) and
(\ref{3.45}) it follows that in this case $\varepsilon=1$.

It remains to consider the case when $c_{y}=0$. Then the second
equation of the system (\ref{3.41}) has the form $1+c_{x}x=0$.
Solving this for $x$ and inserting the result into the first
equation, yields
\begin{equation}
y^{2}=\frac{(e+c_{x}p)^{2}-1}{c_{x}^{2}}
=\frac{\varepsilon^{2}-1}{c_{x}^{2}}.
\label{3.46}
\end{equation}
We see that in this case the conic (\ref{3.9}) and the singular line
of the projective transformation (\ref{3.17}) intersect at two
distinct points for $\varepsilon>1$, have a single point of tangency
for $\varepsilon=1$, and have no points of contact for
$0\leqslant\varepsilon<1$.

\begin{figure}[p]
\centering
\begin{subfigure}[c]{0.475\textwidth}
\centering
\fbox{\includegraphics[width=\linewidth]{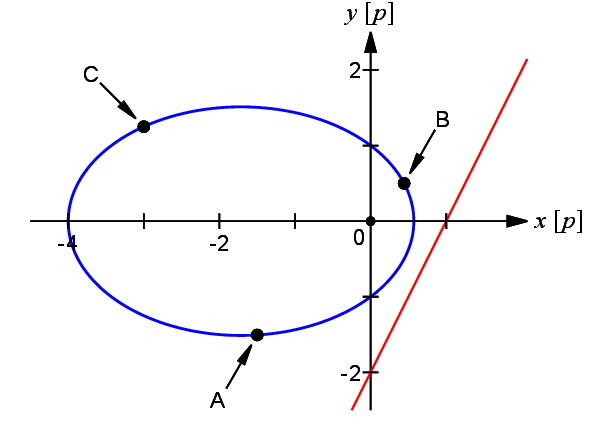}}
\end{subfigure}
\hfill
\begin{subfigure}[c]{0.475\textwidth}
\centering
\fbox{\includegraphics[width=\linewidth]{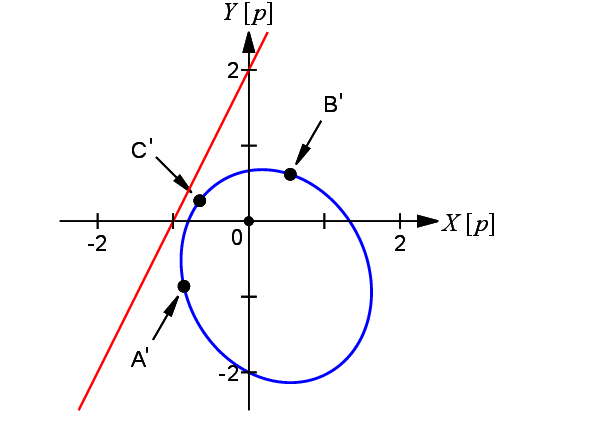}}
\end{subfigure}
\caption{\textbf{Left panel:} The orbit of a particle with energy
$E<0$ is an ellipse with one of its foci at the center of the force,
$(x,y)=(0,0)$. $\mathsf{A}$, $\mathsf{B}$, and $\mathsf{C}$ are three
arbitrary distinct points on this orbit. The singular line
$1+c_{x}x+c_{y}y=0$ of the projective transformation (\ref{3.17}),
also shown, is disjoint from the ellipse. \textbf{Right panel:} The
primed symbols indicate the images of the points marked in the left
panel. The transformed orbit is an ellipse with one of its foci
coinciding with the center of the resulting force at $(X,Y)=(0,0)$.
The line $1-c_{x}X-c_{y}Y=0$, which is the image of the line at
infinity under the transformation (\ref{3.17}), is also indicated. In
both panels, the axes are marked in units of $p$. Parameters:
$e=3/4$, $c_{x}p=-1$, $c_{y}p=1/2$.}
\label{fig1}
\end{figure}

\begin{figure}[p]
\centering
\begin{subfigure}[c]{0.475\textwidth}
\centering
\fbox{\includegraphics[width=\linewidth]{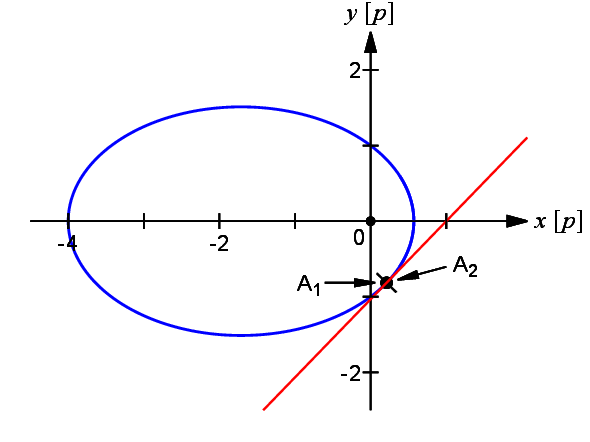}}
\end{subfigure}
\hfill
\begin{subfigure}[c]{0.475\textwidth}
\centering
\fbox{\includegraphics[width=\linewidth]{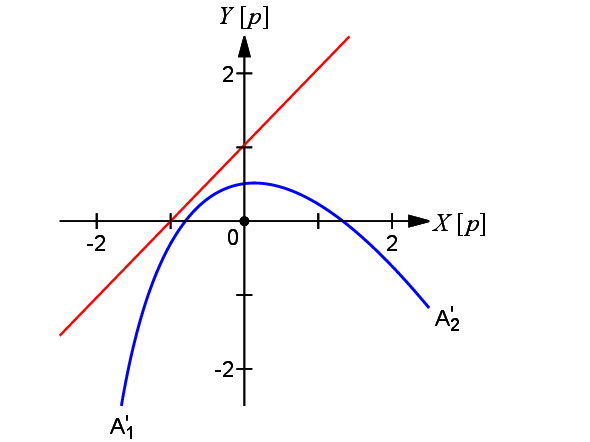}}
\end{subfigure}
\caption{\textbf{Left panel:} As in Fig.\/~\ref{fig1} (left), except
that here the line $1+c_{x}x+c_{y}y=0$ is tangent to the elliptic
orbit. $\mathsf{A_{1}}$ and $\mathsf{A_{2}}$ denote the one-sided
limits at the tangency point, taken along the orbit. \textbf{Right
panel:} The primed symbols indicate the two images, which lie at
infinity, of the limit points marked in the left panel. The
transformed orbit is a parabola with its focus coinciding with the
center of the resulting force at $(X,Y)=(0,0)$. The line
$1-c_{x}X-c_{y}Y=0$ is also indicated. Parameters: $e=3/4$,
$c_{x}p=-1$, $c_{y}p=\sqrt{15}/4$.}
\label{fig2}
\end{figure}

\begin{figure}[p]
\centering
\begin{subfigure}[c]{0.475\textwidth}
\centering
\fbox{\includegraphics[width=\linewidth]{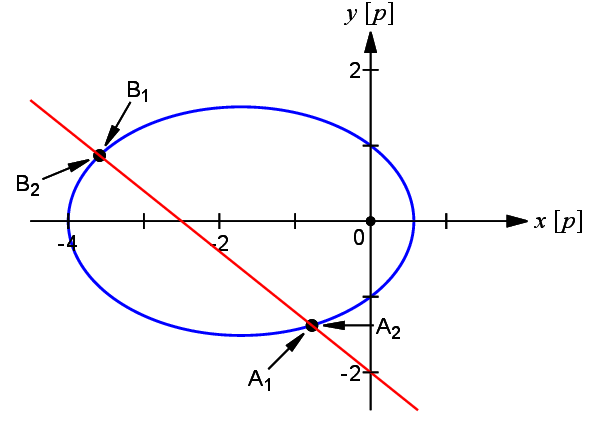}}
\end{subfigure}
\hfill
\begin{subfigure}[c]{0.475\textwidth}
\centering
\fbox{\includegraphics[width=\linewidth]{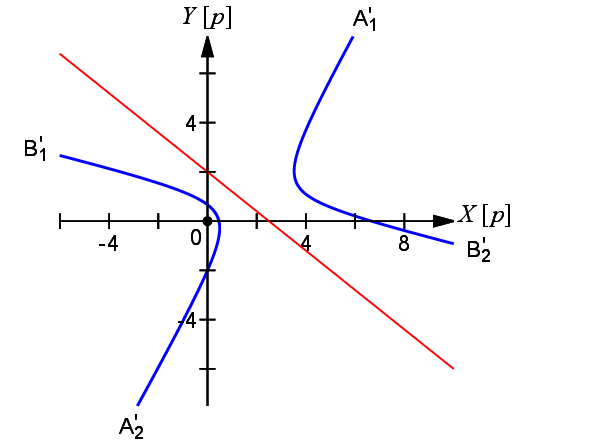}}
\end{subfigure}
\caption{\textbf{Left panel:} As in Fig.\/~\ref{fig1} (left), except
that here the line $1+c_{x}x+c_{y}y=0$ intersects the elliptic orbit
at two distinct points. $\mathsf{A_{1}}$, $\mathsf{A_{2}}$ and
$\mathsf{B_{1}}$, $\mathsf{B_{2}}$ denote the one-sided limits at
these points. \textbf{Right panel:} The primed symbols indicate the
images, all lying at infinity, of the limit points marked in the left
panel. The elliptic arc $\mathsf{B_{2}A_{1}}$ maps onto the repulsive
branch $\mathsf{B_{2}^{\prime}A_{1}^{\prime}}$, whereas the
complementary arc $\mathsf{A_{2}B_{1}}$ maps onto the attractive
branch $\mathsf{A_{2}^{\prime}B_{1}^{\prime}}$, both branches forming
a hyperbola with one of its foci coinciding with the center of the
resulting force at $(X,Y)=(0,0)$. The line $1-c_{x}X-c_{y}Y=0$ is
also indicated. Parameters: $e=3/4$, $c_{x}p=2/5$, $c_{y}p=1/2$.}
\label{fig3}
\end{figure}

\begin{figure}[p]
\centering
\begin{subfigure}[c]{0.475\textwidth}
\centering
\fbox{\includegraphics[width=\linewidth]{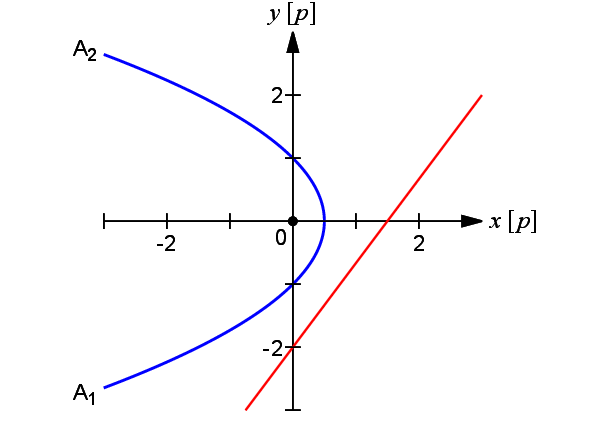}}
\end{subfigure}
\hfill
\begin{subfigure}[c]{0.475\textwidth}
\centering
\fbox{\includegraphics[width=\linewidth]{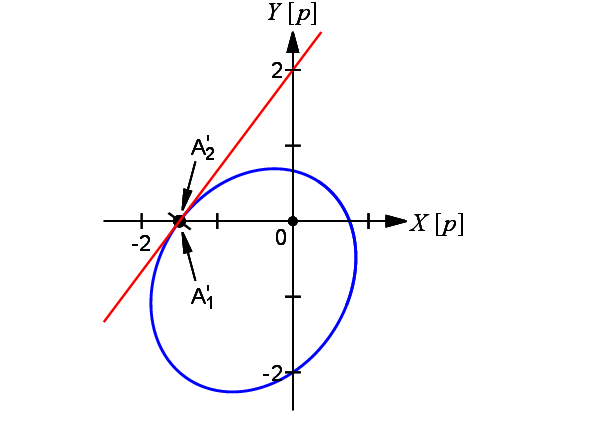}}
\end{subfigure}
\caption{\textbf{Left panel:} The orbit of a particle with energy
$E=0$ is a parabola with its focus at the center of the force,
$(x,y)=(0,0)$. $\mathsf{A_{1}}$ and $\mathsf{A_{2}}$ represent the
points at infinity on the parabola's arms. The singular line
$1+c_{x}x+c_{y}y=0$ of the projective transformation (\ref{3.17}),
also shown, is disjoint from the orbit. \textbf{Right panel:} The
primed symbols indicate the images of the points at infinity marked
in the left panel. The transformed orbit is an ellipse with one of
its foci coinciding with the center of the resulting force at
$(X,Y)=(0,0)$. The line $1-c_{x}X-c_{y}Y=0$, which is the image of
the line at infinity under the transformation (\ref{3.17}), is also
indicated; it is tangent to the ellipse at $(X,Y)=(1/c_{x},0)$ for
any relevant value of $c_{y}$. In both panels, the axes are marked in
units of $p$. Parameters: $e=1$, $c_{x}p=-2/3$, $c_{y}p=1/2$.}
\label{fig4}
\end{figure}

\begin{figure}[p]
\centering
\begin{subfigure}[c]{0.475\textwidth}
\centering
\fbox{\includegraphics[width=\linewidth]{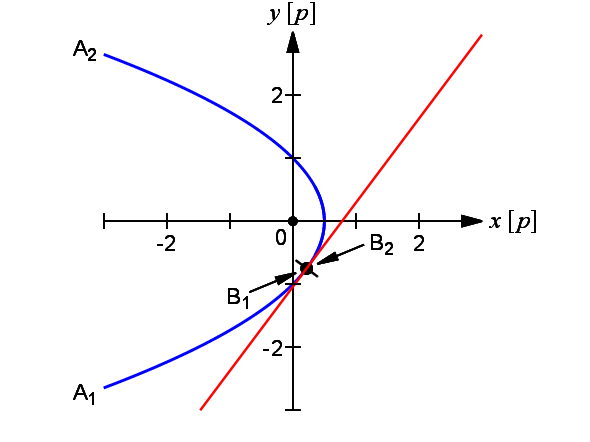}}
\end{subfigure}
\hfill
\begin{subfigure}[c]{0.475\textwidth}
\centering
\fbox{\includegraphics[width=\linewidth]{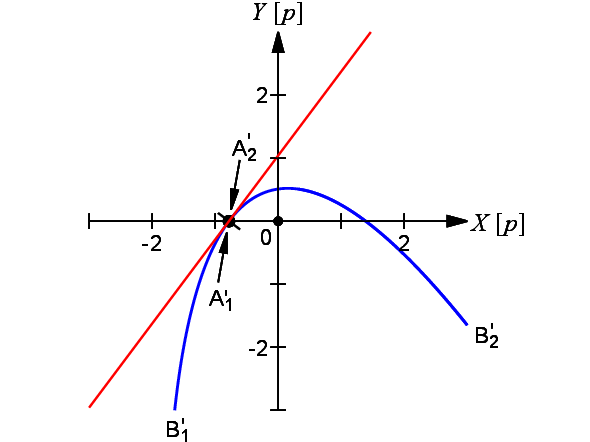}}
\end{subfigure}
\caption{\textbf{Left panel:} As in Fig.\/~\ref{fig4} (left), except
that here the line $1+c_{x}x+c_{y}y=0$ is tangent to the parabolic
orbit. $\mathsf{B_{1}}$ and $\mathsf{B_{2}}$ denote the one-sided
limits at the tangency point, taken along the orbit. \textbf{Right
panel:} The primed symbols indicate the images of the limit points
marked in the left panel. The transformed orbit is a parabola with its
focus coinciding with the center of the resulting force at
$(X,Y)=(0,0)$. The line $1-c_{x}X-c_{y}Y=0$ is also indicated; it is
tangent to the image parabola at $(X,Y)=(1/c_{x},0)$ for any relevant
value of $c_{y}$. Parameters: $e=1$, $c_{x}p=-32/25$, $c_{y}p=24/25$.}
\label{fig5}
\end{figure}

\begin{figure}[p]
\centering
\begin{subfigure}[c]{0.475\textwidth}
\centering
\fbox{\includegraphics[width=\linewidth]{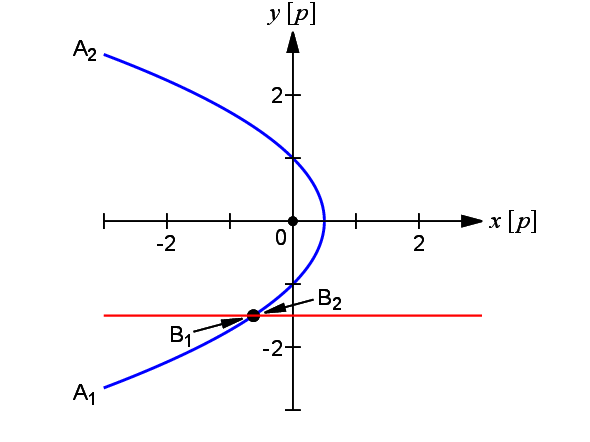}}
\end{subfigure}
\hfill
\begin{subfigure}[c]{0.475\textwidth}
\centering
\fbox{\includegraphics[width=\linewidth]{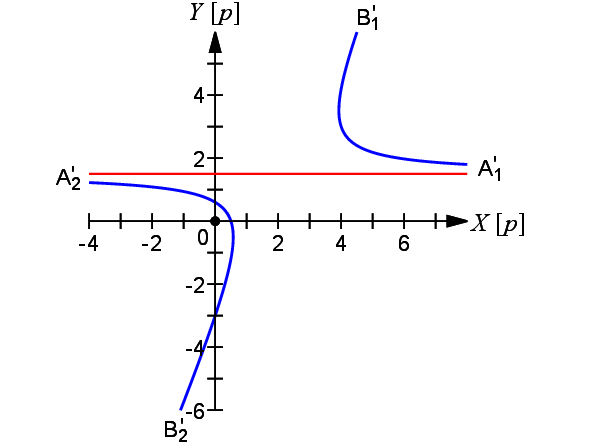}}
\end{subfigure}
\caption{\textbf{Left panel:} As in Fig.\/~\ref{fig4} (left), except
that here the line $1+c_{x}x+c_{y}y=0$ is parallel to the parabola's
axis of symmetry (hence, $c_{x}=0$) and intersects the orbit at a
single point. $\mathsf{B_{1}}$ and $\mathsf{B_{2}}$ denote the
one-sided limits at the intersection point, taken along the orbit.
\textbf{Right panel:} The primed symbols indicate the images of the
limit points marked in the left panel. The semi-infinite parabolic
arc $\mathsf{A_{1}B_{1}}$ maps onto the repulsive branch
$\mathsf{A_{1}^{\prime}B_{1}^{\prime}}$, whereas the complementary
parabolic arc $\mathsf{B_{2}A_{2}}$ maps onto the attractive branch
$\mathsf{B_{2}^{\prime}A_{2}^{\prime}}$, the two branches forming a
hyperbola with one of its foci coinciding with the center of the
resulting force at $(X,Y)=(0,0)$. The line $1-c_{y}Y=0$, also
indicated, is an asymptote of the image hyperbola. Parameters: $e=1$,
$c_{x}p=0$ (cf.\ above), $c_{y}p=2/3$.}
\label{fig6}
\end{figure}

\begin{figure}[p]
\centering
\begin{subfigure}[c]{0.475\textwidth}
\centering
\fbox{\includegraphics[width=\linewidth]{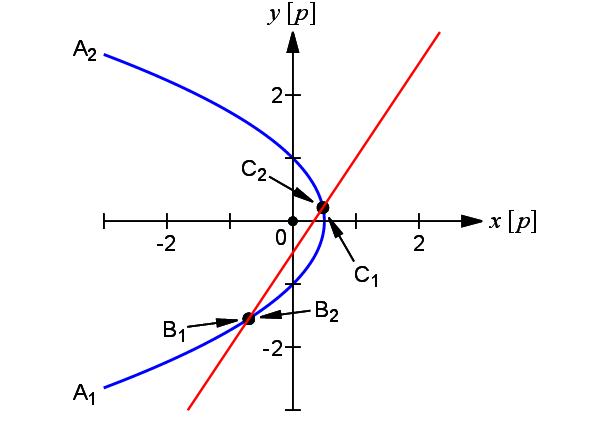}}
\end{subfigure}
\hfill
\begin{subfigure}[c]{0.475\textwidth}
\centering
\fbox{\includegraphics[width=\linewidth]{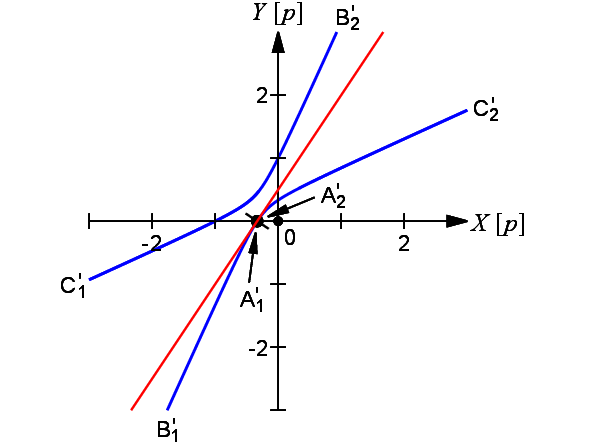}}
\end{subfigure}
\caption{\textbf{Left panel:} As in Fig.\/~\ref{fig4} (left), except
that here the line $1+c_{x}x+c_{y}y=0$ intersects the parabolic orbit
at two distinct points. $\mathsf{B_{1}}$, $\mathsf{B_{2}}$ and
$\mathsf{C_{1}}$, $\mathsf{C_{2}}$ denote the one-sided limits at
these intersection points, taken along the orbit. \textbf{Right
panel:} The primed symbols indicate the images of the limit points
marked in the left panel. The two semi-infinite parabolic arcs
$\mathsf{A_{1}B_{1}}$ and $\mathsf{C_{2}A_{2}}$ map, respectively,
onto the semi-infinite arcs $\mathsf{A_{1}^{\prime}B_{1}^{\prime}}$
and $\mathsf{C_{2}^{\prime}A_{2}^{\prime}}$. Together, the two image
arcs form an attractive branch of a hyperbola with one of its foci
coinciding with the center of the resulting force at $(X,Y)=(0,0)$.
The finite parabolic arc $\mathsf{B_{2}C_{1}}$ maps onto the entire
repulsive branch $\mathsf{B_{2}^{\prime}C_{1}^{\prime}}$ of the same
hyperbola. The line $1-c_{x}X-c_{y}Y=0$ is also indicated; it is
tangent to the attractive hyperbolic branch at $(X,Y)=(1/c_{x},0)$
for any relevant value of $c_{y}$. Parameters: $e=1$, $c_{x}p=-3$,
$c_{y}p=2$.}
\label{fig7}
\end{figure}

\begin{figure}[p]
\centering
\begin{subfigure}[c]{0.475\textwidth}
\centering
\fbox{\includegraphics[width=\linewidth]{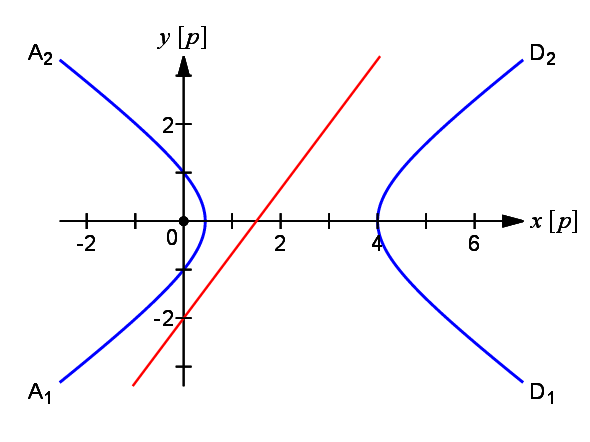}}
\end{subfigure}
\hfill
\begin{subfigure}[c]{0.475\textwidth}
\centering
\fbox{\includegraphics[width=\linewidth]{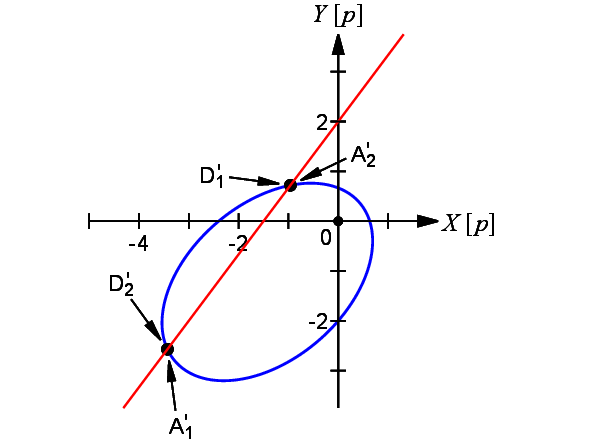}}
\end{subfigure}
\caption{\textbf{Left panel:} The orbit of a particle with energy
$E>0$ is one of the two branches of a hyperbola with one of its foci
at the center of the force, $(x,y)=(0,0)$. $\mathsf{A_{1}}$,
$\mathsf{A_{2}}$ and $\mathsf{D_{1}}$, $\mathsf{D_{2}}$ denote the
points at infinity along the asymptotic directions of the respective
branches. The branch $\mathsf{A_{1}A_{2}}$ corresponds to motion in
an attractive field ($\alpha<0$), whereas the branch
$\mathsf{D_{1}D_{2}}$ corresponds to motion in a repulsive field
($\alpha>0$). The singular line $1+c_{x}x+c_{y}y=0$ of the
projective transformation (\ref{3.17}), also shown, does not
intersect either branch. \textbf{Right panel:} The primed symbols
indicate the images of the limit points at infinity marked in the
left panel. The attractive orbit $\mathsf{A_{1}A_{2}}$ maps onto the
arc $\mathsf{A_{1}^{\prime}A_{2}^{\prime}}$ of an ellipse with one of
its foci coinciding with the center of the resulting force at
$(X,Y)=(0,0)$. The repulsive orbit $\mathsf{D_{1}D_{2}}$ maps onto
the complementary elliptic arc
$\mathsf{D_{1}^{\prime}D_{2}^{\prime}}$. The line
$1-c_{x}X-c_{y}Y=0$, which is the image of the line at infinity under
the transformation (\ref{3.17}), is also indicated. In both panels,
the axes are marked in units of $p$. Parameters: $e=5/4$,
$c_{x}p=-2/3$, $c_{y}p=1/2$.}
\label{fig8}
\end{figure}

\begin{figure}[p]
\centering
\begin{subfigure}[c]{0.475\textwidth}
\centering
\fbox{\includegraphics[width=\linewidth]{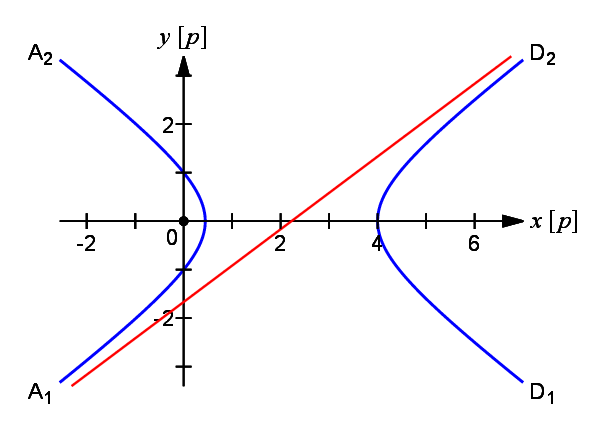}}
\end{subfigure}
\hfill
\begin{subfigure}[c]{0.475\textwidth}
\centering
\fbox{\includegraphics[width=\linewidth]{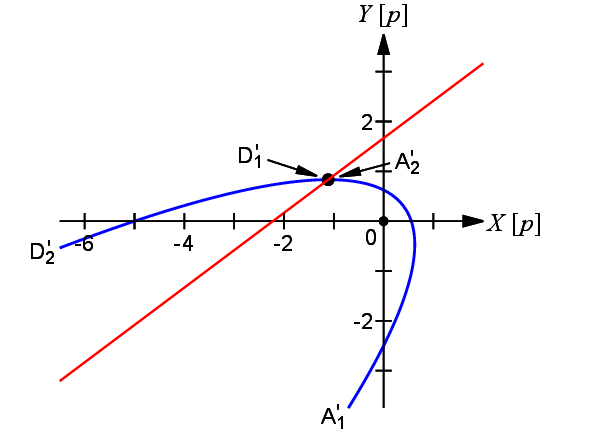}}
\end{subfigure}
\caption{\textbf{Left panel:} As in Fig.\/~\ref{fig8} (left), except
that here the line $1+c_{x}x+c_{y}y=0$, with
\mbox{$c_{x}=-(e^{2}-1)/(ep)$} and $c_{y}=\sqrt{e^{2}-1}/(ep)$,
coincides with the hyperbola's asymptote
$y=\sqrt{e^{2}-1}\,x-ep/\sqrt{e^{2}-1}$. \textbf{Right panel:} The
primed symbols indicate the images of the limit points marked in the
left panel. The attractive orbit $\mathsf{A_{1}A_{2}}$ maps onto the
semi-infinite arc $\mathsf{A_{1}^{\prime}A_{2}^{\prime}}$ of a
parabola with its focus coinciding with the center of the resulting
force at $(X,Y)=(0,0)$. The repulsive orbit $\mathsf{D_{1}D_{2}}$
maps onto the complementary semi-infinite parabolic arc
$\mathsf{D_{1}^{\prime}D_{2}^{\prime}}$. The line $1-c_{x}X-c_{y}Y=0$
is also indicated. Parameters: $e=5/4$, $c_{x}p=-9/20$, $c_{y}p=3/5$.}
\label{fig9}
\end{figure}

\begin{figure}[p]
\centering
\begin{subfigure}[c]{0.475\textwidth}
\centering
\fbox{\includegraphics[width=\linewidth]{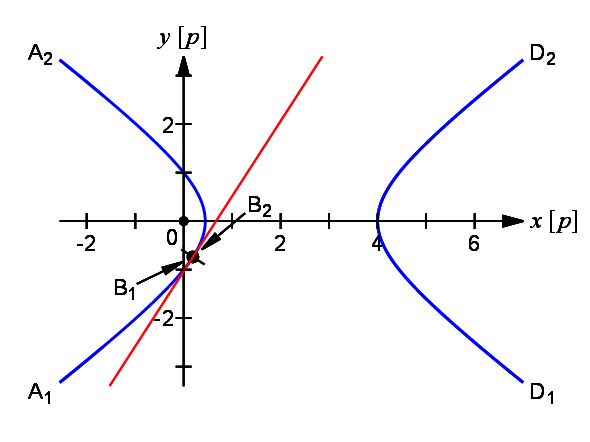}}
\end{subfigure}
\hfill
\begin{subfigure}[c]{0.475\textwidth}
\centering
\fbox{\includegraphics[width=\linewidth]{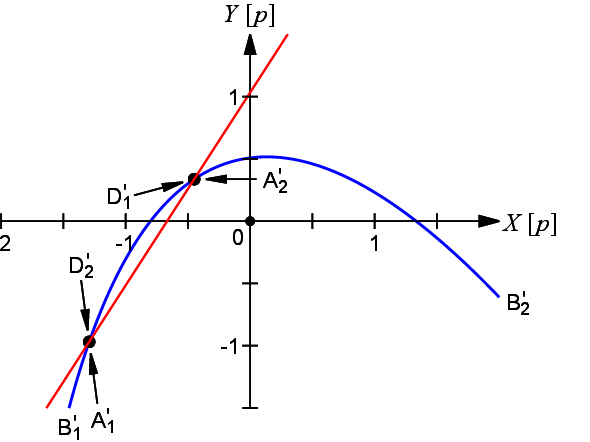}}
\end{subfigure}
\caption{\textbf{Left panel:} As in Fig.\/~\ref{fig8} (left), except
that here the line $1+c_{x}x+c_{y}y=0$ is tangent to the
attractive orbit and has no points in common with the partner
repulsive orbit. $\mathsf{B_{1}}$ and $\mathsf{B_{2}}$ denote the
one-sided limits at the tangency point, taken along the orbit.
\textbf{Right panel:} The primed symbols indicate the images of the
limit points marked in the left panel. The attractive orbit
$\mathsf{A_{1}A_{2}}$ maps onto the union of two semi-infinite arcs
$\mathsf{A_{1}^{\prime}B_{1}^{\prime}}$ and
$\mathsf{B_{2}^{\prime}A_{2}^{\prime}}$ of a parabola with its focus
coinciding with the center of the resulting force at $(X,Y)=(0,0)$.
The repulsive orbit $\mathsf{D_{1}D_{2}}$ maps onto the complementary
finite parabolic arc $\mathsf{D_{1}^{\prime}D_{2}^{\prime}}$. The
line $1-c_{x}X-c_{y}Y=0$ is also indicated. Parameters: $e=5/4$,
$c_{x}p=-3/2$, $c_{y}p=\sqrt{15}/4$.}
\label{fig10}
\end{figure}

\begin{figure}[p]
\centering
\begin{subfigure}[c]{0.475\textwidth}
\centering
\fbox{\includegraphics[width=\linewidth]{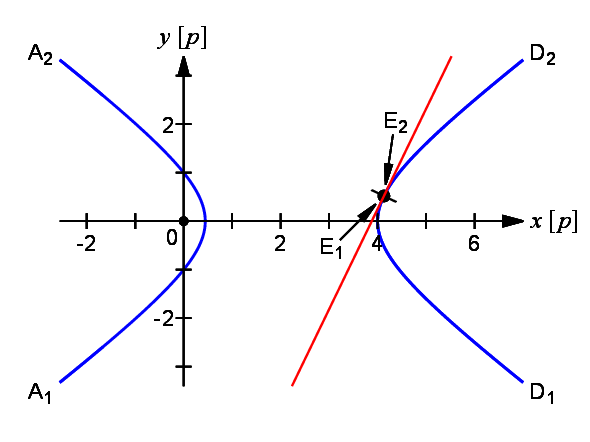}}
\end{subfigure}
\hfill
\begin{subfigure}[c]{0.475\textwidth}
\centering
\fbox{\includegraphics[width=\linewidth]{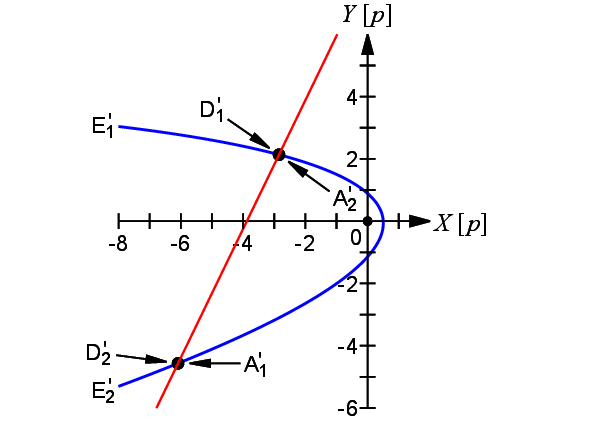}}
\end{subfigure}
\caption{\textbf{Left panel:} As in Fig.\/~\ref{fig8} (left), except
that here the line $1+c_{x}x+c_{y}y=0$ is tangent to the repulsive
orbit and has no points in common with the partner attractive orbit.
$\mathsf{E_{1}}$ and $\mathsf{E_{2}}$ denote the one-sided limits at
the tangency point, taken along the orbit. \textbf{Right panel:} The
primed symbols indicate the images of the limit points marked in the
left panel. The attractive orbit $\mathsf{A_{1}A_{2}}$ maps onto the
finite arc $\mathsf{A_{1}^{\prime}A_{2}^{\prime}}$ of a parabola with
its focus coinciding with the center of the resulting force at
$(X,Y)=(0,0)$. The repulsive orbit $\mathsf{D_{1}D_{2}}$ maps onto
the union of two complementary semi-infinite parabolic arcs
$\mathsf{D_{1}^{\prime}E_{1}^{\prime}}$ and
$\mathsf{E_{2}^{\prime}D_{2}^{\prime}}$. The line $1-c_{x}X-c_{y}Y=0$
is also indicated. Parameters: $e=5/4$, $c_{x}p=-33/128$,
$c_{y}p=\sqrt{255}/128$.}
\label{fig11}
\end{figure}

\begin{figure}[p]
\centering
\begin{subfigure}[c]{0.475\textwidth}
\centering
\fbox{\includegraphics[width=\linewidth]{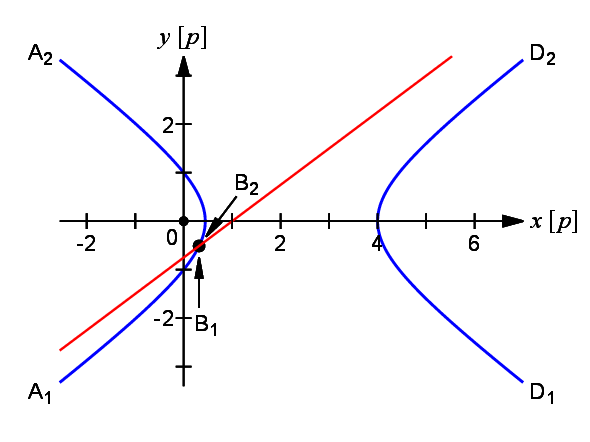}}
\end{subfigure}
\hfill
\begin{subfigure}[c]{0.475\textwidth}
\centering
\fbox{\includegraphics[width=\linewidth]{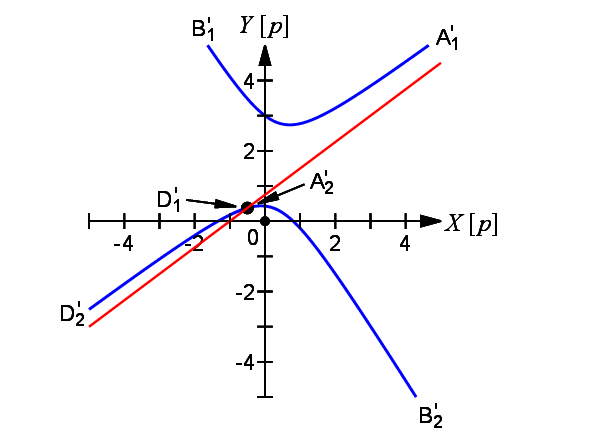}}
\end{subfigure}
\caption{\textbf{Left panel:} As in Fig.\/~\ref{fig8} (left), except
that here the line $1+c_{x}x+c_{y}y=0$, with
$c_{x}/c_{y}=-\sqrt{e^{2}-1}$, is parallel to, but does not lie
along, the hyperbola's asymptote
$y=\sqrt{e^{2}-1}\,x-ep/\sqrt{e^{2}-1}$; it intersects the
attractive orbit at a single point and has no points in common with
the partner repulsive orbit. $\mathsf{B_{1}}$ and $\mathsf{B_{2}}$
denote the one-sided limits at the intersection point, taken along
the orbit. \textbf{Right panel:} The primed symbols indicate the
images of the limit points marked in the left panel. The attractive
orbit $\mathsf{A_{1}A_{2}}$ maps onto the union of the repulsive
branch $\mathsf{A_{1}^{\prime}B_{1}^{\prime}}$ and the semi-infinite
attractive arc $\mathsf{B_{2}^{\prime}A_{2}^{\prime}}$, both
belonging to a hyperbola with one of its foci coinciding with the
center of the resulting force at $(X,Y)=(0,0)$. The repulsive orbit
$\mathsf{D_{1}D_{2}}$ maps onto the semi-infinite attractive
hyperbolic arc $\mathsf{D_{1}^{\prime}D_{2}^{\prime}}$ complementary
to $\mathsf{B_{2}^{\prime}A_{2}^{\prime}}$. The line
$1-c_{x}X-c_{y}Y=0$ is also indicated. Parameters: $e=5/4$,
$c_{x}p=-1$, $c_{y}p=4/3$.}
\label{fig12}
\end{figure}

\begin{figure}[p]
\centering
\begin{subfigure}[c]{0.475\textwidth}
\centering
\fbox{\includegraphics[width=\linewidth]{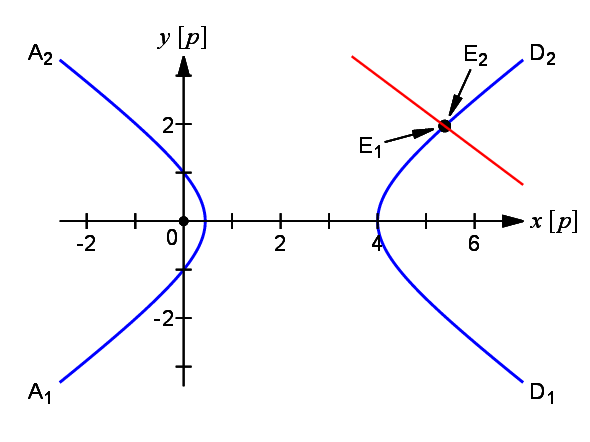}}
\end{subfigure}
\hfill
\begin{subfigure}[c]{0.475\textwidth}
\centering
\fbox{\includegraphics[width=\linewidth]{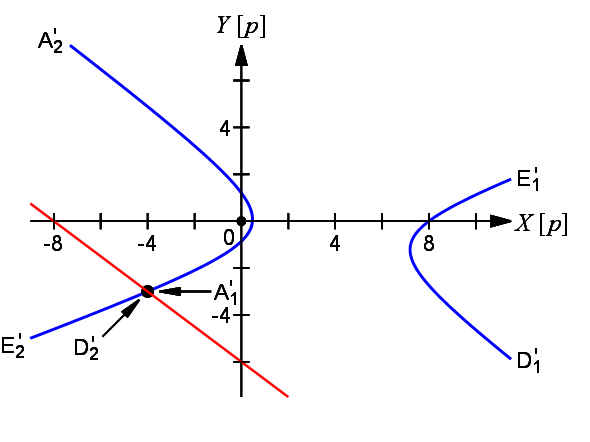}}
\end{subfigure}
\caption{\textbf{Left panel:} As in Fig.\/~\ref{fig8} (left), except
that here the line $1+c_{x}x+c_{y}y=0$, with
$c_{x}/c_{y}=\sqrt{e^{2}-1}$, is parallel to, but does not lie along,
the hyperbola's asymptote $y=-\sqrt{e^{2}-1}\,x+ep/\sqrt{e^{2}-1}$;
it intersects the repulsive orbit at a single point and has no points
in common with the partner attractive orbit. $\mathsf{E_{1}}$ and
$\mathsf{E_{2}}$ denote the one-sided limits at the intersection
point, taken along the orbit. \textbf{Right panel:} The primed
symbols indicate the images of the limit points marked in the left
panel. The attractive orbit $\mathsf{A_{1}A_{2}}$ maps onto the
semi-infinite attractive arc $\mathsf{A_{1}^{\prime}A_{2}^{\prime}}$
of a hyperbola with one of its foci coinciding with the center of the
resulting force at $(X,Y)=(0,0)$. The repulsive orbit
$\mathsf{D_{1}D_{2}}$ maps onto the union of the repulsive hyperbolic
branch $\mathsf{D_{1}^{\prime}E_{1}^{\prime}}$ and the semi-infinite
attractive hyperbolic arc $\mathsf{E_{2}^{\prime}D_{2}^{\prime}}$
complementary to $\mathsf{A_{1}^{\prime}A_{2}^{\prime}}$. The line
$1-c_{x}X-c_{y}Y=0$ is also indicated. Parameters: $e=5/4$,
$c_{x}p=-1/8$, $c_{y}p=-1/6$.}
\label{fig13}
\end{figure}

\begin{figure}[p]
\centering
\begin{subfigure}[c]{0.475\textwidth}
\centering
\fbox{\includegraphics[width=\linewidth]{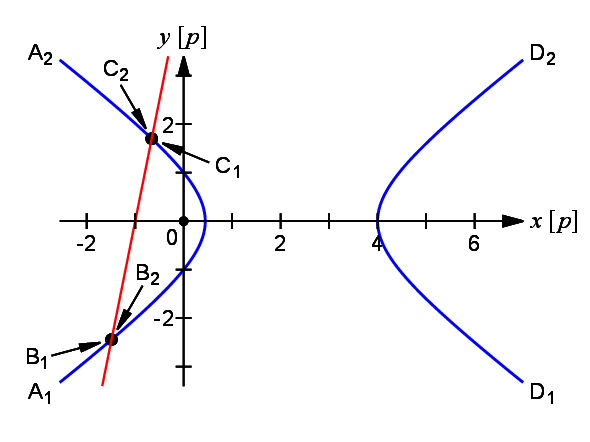}}
\end{subfigure}
\hfill
\begin{subfigure}[c]{0.475\textwidth}
\centering
\fbox{\includegraphics[width=\linewidth]{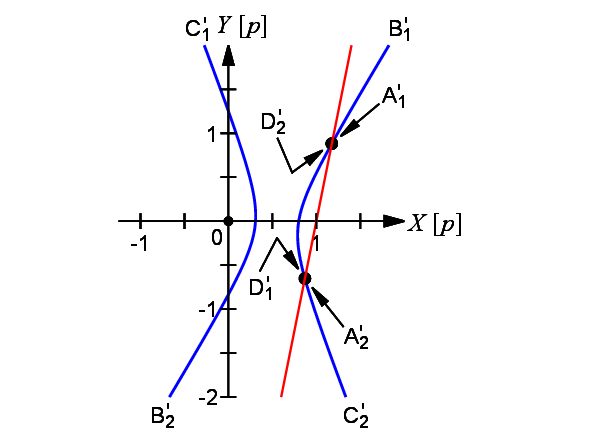}}
\end{subfigure}
\caption{\textbf{Left panel:} As in Fig.\/~\ref{fig8} (left), except
that here the line $1+c_{x}x+c_{y}y=0$ intersects the attractive
orbit at two distinct points and has no points in common with the
partner repulsive orbit. $\mathsf{B_{1}}$, $\mathsf{B_{2}}$ and
$\mathsf{C_{1}}$, $\mathsf{C_{2}}$ denote the one-sided limits at the
respective intersection points, taken along the orbit. \textbf{Right
panel:} The primed symbols indicate the images of the limit points
marked in the left panel. The attractive orbit $\mathsf{A_{1}A_{2}}$
maps onto the union of the semi-infinite repulsive arc
$\mathsf{A_{1}^{\prime}B_{1}^{\prime}}$, the attractive branch
$\mathsf{B_{2}^{\prime}C_{1}^{\prime}}$, and the semi-infinite
repulsive arc $\mathsf{C_{2}^{\prime}A_{2}^{\prime}}$, all belonging
to a hyperbola with one of its foci coinciding with the center of the
resulting force at $(X,Y)=(0,0)$. The repulsive orbit
$\mathsf{D_{1}D_{2}}$ maps onto the finite repulsive hyperbolic arc
$\mathsf{D_{1}^{\prime}D_{2}^{\prime}}$, complementary to
$\mathsf{C_{2}^{\prime}A_{2}^{\prime}}$ and
$\mathsf{A_{1}^{\prime}B_{1}^{\prime}}$. The line $1-c_{x}X-c_{y}Y=0$
is also indicated. Parameters: $e=5/4$, $c_{x}p=1$, $c_{y}p=-1/5$.}
\label{fig14}
\end{figure}

\begin{figure}[p]
\centering
\begin{subfigure}[c]{0.475\textwidth}
\centering
\fbox{\includegraphics[width=\linewidth]{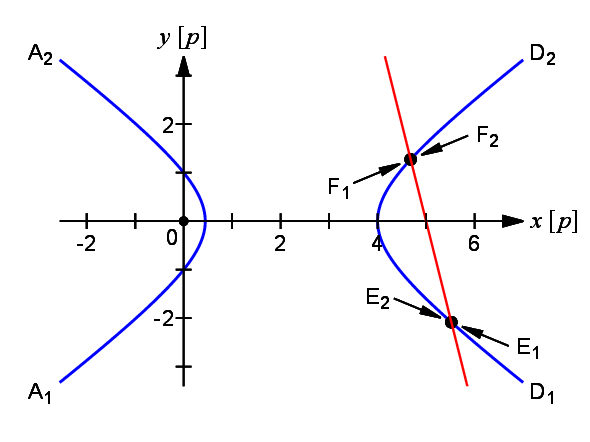}}
\end{subfigure}
\hfill
\begin{subfigure}[c]{0.475\textwidth}
\centering
\fbox{\includegraphics[width=\linewidth]{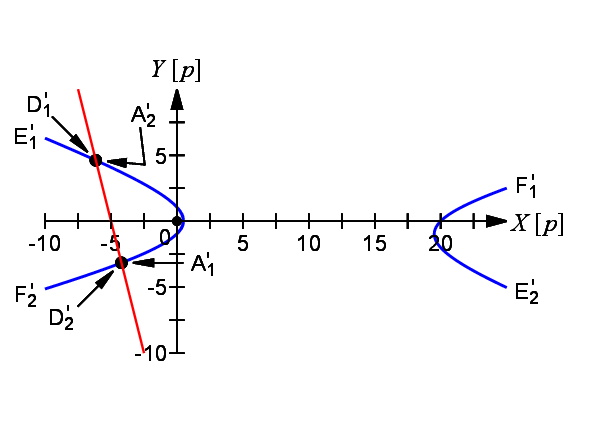}}
\end{subfigure}
\caption{\textbf{Left panel:} As in Fig.\/~\ref{fig8} (left), except
that here the line $1+c_{x}x+c_{y}y=0$ intersects the repulsive orbit
at two distinct points and has no points in common with the partner
attractive orbit. $\mathsf{E_{1}}$, $\mathsf{E_{2}}$ and
$\mathsf{F_{1}}$, $\mathsf{F_{2}}$ denote the one-sided limits at the
respective intersection points, taken along the orbit. \textbf{Right
panel:} The primed symbols indicate the images of the limit points
marked in the left panel. The attractive orbit $\mathsf{A_{1}A_{2}}$
maps onto the finite attractive arc
$\mathsf{A_{1}^{\prime}A_{2}^{\prime}}$ of a hyperbola with one of
its foci coinciding with the center of the resulting force at
$(X,Y)=(0,0)$. The repulsive orbit $\mathsf{D_{1}D_{2}}$ maps onto the
union of the semi-infinite attractive hyperbolic arc
$\mathsf{D_{1}^{\prime}E_{1}^{\prime}}$, the repulsive hyperbolic
branch $\mathsf{E_{2}^{\prime}F_{1}^{\prime}}$, and the semi-infinite
attractive hyperbolic arc $\mathsf{F_{2}^{\prime}D_{2}^{\prime}}$,
the two arcs being complementary to
$\mathsf{A_{1}^{\prime}A_{2}^{\prime}}$. The line $1-c_{x}X-c_{y}Y=0$
is also indicated. Parameters: $e=5/4$, $c_{x}p=-1/5$,
$c_{y}p=-1/20$.}
\label{fig15}
\end{figure}

\begin{figure}[p]
\centering
\begin{subfigure}[c]{0.475\textwidth}
\centering
\fbox{\includegraphics[width=\linewidth]{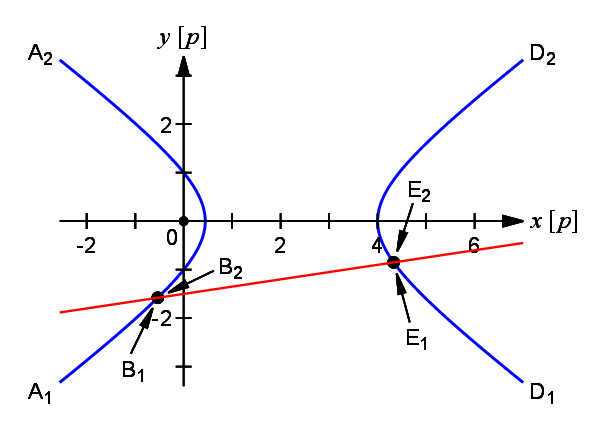}}
\end{subfigure}
\hfill
\begin{subfigure}[c]{0.475\textwidth}
\centering
\fbox{\includegraphics[width=\linewidth]{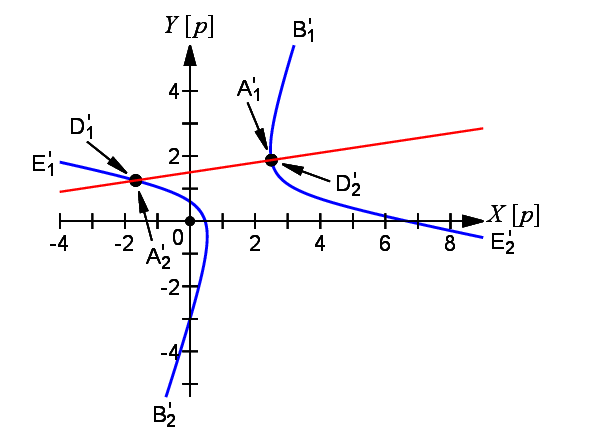}}
\end{subfigure}
\caption{\textbf{Left panel:} As in Fig.\/~\ref{fig8} (left), except
that here the line $1+c_{x}x+c_{y}y=0$ intersects each of the two
partner hyperbolic orbits at a single point. $\mathsf{B_{1}}$,
$\mathsf{B_{2}}$, and $\mathsf{E_{1}}$, $\mathsf{E_{2}}$ denote the
one-sided limits at the respective intersection points, taken along
the orbit. \textbf{Right panel:} The primed symbols indicate the
images of the limit points marked in the left panel. The attractive
orbit $\mathsf{A_{1}A_{2}}$ maps onto the union of the semi-infinite
repulsive arc $\mathsf{A_{1}^{\prime}B_{1}^{\prime}}$ and the
semi-infinite attractive arc $\mathsf{B_{2}^{\prime}A_{2}^{\prime}}$,
both belonging to a hyperbola with one of its foci coinciding with
the center of the resulting force at $(X,Y)=(0,0)$. The repulsive
orbit $\mathsf{D_{1}D_{2}}$ maps onto the union of two hyperbolic
arcs: the semi-infinite attractive arc
$\mathsf{D_{1}^{\prime}E_{1}^{\prime}}$, complementary to
$\mathsf{B_{2}^{\prime}A_{2}^{\prime}}$, and the semi-infinite
repulsive arc $\mathsf{E_{2}^{\prime}D_{2}^{\prime}}$, complementary
to $\mathsf{A_{1}^{\prime}B_{1}^{\prime}}$. The line
$1-c_{x}X-c_{y}Y=0$ is also indicated. Parameters: $e=5/4$,
$c_{x}p=-1/10$, $c_{y}p=2/3$.}
\label{fig16}
\end{figure}

The above analysis leads to the following conclusions, illustrated in
Figs.\/~\ref{fig1}--\ref{fig16}:
\begin{list}{}{}
\item[1.] If the line $1+c_{x}x+c_{y}y=0$ has no points of contact
with the particle's orbit (nor with the partner hyperbolic orbit when
$e>1$), and if it does not coincide with either asymptote of the
orbit when $e>1$, then $0\leqslant\varepsilon<1$ and the transformed
curve is an ellipse or a subset of it; cf.\ Figs.\/~\ref{fig1},
\ref{fig4}, and \ref{fig8}.
\item[2a.] If $e>1$ and if the line $1+c_{x}x+c_{y}y=0$ coincides
with either asymptote of the particle's hyperbolic orbit, then
$\varepsilon=1$ and the transformed curve is a subset of a parabola
(the partner hyperbolic orbit maps onto a complementary arc of that
parabola); cf.\ Fig.\/~\ref{fig9}.
\item[2b.] If the line $1+c_{x}x+c_{y}y=0$ has a point of tangency
with the particle's orbit (or with the partner hyperbolic orbit when
$e>1$), then $\varepsilon=1$ and the transformed curve is a parabola
or a subset of it; cf.\ Figs.\/~\ref{fig2}, \ref{fig5}, \ref{fig10},
and \ref{fig11}.
\item[3a.] If $e\geqslant1$ and if the line $1+c_{x}x+c_{y}y=0$
intersects the particle's orbit (or the partner hyperbolic orbit when
$e>1$) transversely at exactly one point, then $\varepsilon>1$ and
the transformed curve is a hyperbola or a subset of it; cf.\
Figs.\/~\ref{fig6}, \ref{fig12}, and \ref{fig13}.
\item[3b.] If the line $1+c_{x}x+c_{y}y=0$ intersects the particle's
orbit (or the partner hyperbolic orbit when $e>1$) transversely at
two distinct points, then $\varepsilon>1$ and the transformed curve
is a hyperbola or a subset of it; cf.\ Figs.\/~\ref{fig3},
\ref{fig7}, \ref{fig14}, and \ref{fig15}.
\item[3c.] If $e>1$ and if the line $1+c_{x}x+c_{y}y=0$ intersects
transversely both the particle's hyperbolic orbit and the partner
hyperbolic orbit, each at exactly one point, then $\varepsilon>1$ and
the transformed curve is a subset of a hyperbola; cf.\
Fig.\/~\ref{fig16}.
\end{list}
%
%
%
%
%
%
%
%
%
%
%
%
%
%
%
%
%
%
%
%
%
%
%
%
%
%
%
%
%
%
%
%
%
%
\subsection{Mapping of the Kepler--Coulomb velocity hodographs}
\label{III.3}
Compared with the mapping of orbits discussed above, the mapping of
the velocity hodographs is considerably simpler to analyse. Inverting
Eq.\ (\ref{2.15}), we obtain
\begin{equation}
\boldsymbol{V}=\boldsymbol{v}
-\frac{\hat{\mathbf{L}}\cdot\boldsymbol{c}}{m}.
\label{3.47}
\end{equation}
It is seen that the hodograph of $\boldsymbol{V}$ is obtained from
that of $\boldsymbol{v}$ simply through a translation by the constant
vector $-\hat{\mathbf{L}}\cdot\boldsymbol{c}/m$. Consequently, the
geometric classification of admissible hodographs of $\boldsymbol{V}$
coincides with that of $\boldsymbol{v}$ (see the remarks at the end
of Sec.\ \ref{III.1}).

From Eq.\ (\ref{3.47}), the Cartesian components of $\boldsymbol{V}$
in the plane of transformed motion read
\begin{equation}
V_{X}=v_{x}-\frac{c_{y}L}{m},
\qquad
V_{Y}=v_{y}+\frac{c_{x}L}{m}.
\label{3.48}
\end{equation}
Combining this with Eq.\ (\ref{3.15}) results in the parametric
Cartesian representation of the hodograph of $\boldsymbol{V}$ in the
form
\begin{subequations}
\begin{equation}
V_{X}=-\frac{c_{y}L}{m}+\frac{\alpha}{L}\sin\varphi,
\label{3.49a}
\end{equation}
\begin{equation}
V_{Y}=\left(\frac{e|\alpha|}{L}
+\frac{c_{x}L}{m}\right)-\frac{\alpha}{L}\cos\varphi,
\label{3.49b}
\end{equation}
\label{3.49}%
\end{subequations}
with $\varphi$ varying as given in Eq.\ (\ref{3.6}). Eliminating the
parameter $\varphi$ from Eqs.\ (\ref{3.49}) yields the implicit
equation
\begin{equation}
\left(V_{X}+\frac{c_{y}L}{m}\right)^{2}
+\left[V_{Y}-\left(\frac{e|\alpha|}{L}
+\frac{c_{x}L}{m}\right)\right]^{2}=\frac{\alpha^{2}}{L^{2}},
\label{3.50}
\end{equation}
which represents a circle centered at
$(V_{X},V_{Y})=(-c_{y}L/m,e|\alpha|/L+c_{x}L/m)$ with radius
$|\alpha|/|L|$; it is an underlying circle of the hodograph of
$\boldsymbol{V}$.
%
%
\subsection{Relation to the work of Kholshevnikov and Titov}
\label{III.4}
Before concluding this part of our discussion, we should mention the
study by Kholshevnikov and Titov (2007, Sec.\/~2.5), which bears some
relation to the present work, although the approach adopted there is
essentially the reverse of ours. Below, we summarize the reasoning of
these authors in a notation different from the original one but
consistent with that used throughout the present paper.

Kholshevnikov and Titov consider the Kepler case ($\alpha<0$) only
and take advantage of the planarity of motion from the outset. They
begin by translating the particle's velocity hodograph by the
constant vector $\boldsymbol{u}=(u_{x},u_{y})$:
\begin{equation}
V_{X}=v_{x}+u_{x},
\qquad
V_{Y}=v_{y}+u_{y}.
\label{3.51}
\end{equation}
Then, after introducing an additional requirement that the particle's
angular momentum be invariant under this translation,
\begin{equation}
m(XV_{Y}-YV_{X})=m(xv_{y}-yv_{x})
\qquad (=L),
\label{3.52}
\end{equation}
they conclude, by an argument whose details are not entirely clear to
us, that this implies the following transformations of components of
the particle's position vector:
\begin{subequations}
\begin{equation}
X=x\frac{xv_{y}-yv_{x}}{xv_{y}-yv_{x}+xu_{y}-yu_{x}},
\label{3.53a}
\end{equation}
\begin{equation}
Y=y\frac{xv_{y}-yv_{x}}{xv_{y}-yv_{x}+xu_{y}-yu_{x}},
\label{3.53b}
\end{equation}
\label{3.53}%
\end{subequations}
along with the differential time scaling
\begin{equation}
\mathrm{d}T=\left(\frac{\sqrt{m|\alpha|p}}
{\sqrt{m|\alpha|p}+xu_{y}-yu_{x}}\right)^{2}\mathrm{d}t.
\label{3.54}
\end{equation}
Here $p$ is the orbital parameter defined in Eq.\ (\ref{3.2}). The
mathematical reasoning of the authors does not go beyond this point.
However, it is easy to verify that Eqs.\ (\ref{3.53}) and
(\ref{3.54}) can be rewritten in vector-tensor notation as
\begin{equation}
\boldsymbol{R}=\frac{\boldsymbol{r}}
{\displaystyle1+\frac{m\boldsymbol{r}\cdot\hat{\mathbf{L}}
\cdot\boldsymbol{u}}{L^{2}}}
\label{3.55}
\end{equation}
and
\begin{equation}
\mathrm{d}T=\frac{\mathrm{d}t}
{\left(\displaystyle1+\frac{m\boldsymbol{r}\cdot\hat{\mathbf{L}}
\cdot\boldsymbol{u}}{L^{2}}\right)^{2}},
\label{3.56}
\end{equation}
respectively, since in two spatial dimensions
\begin{equation}
\hat{\mathbf{L}}\cdot\hat{\mathbf{L}}=-L^{2}\hat{\mathbf{I}},
\label{3.57}
\end{equation}
where $\hat{\mathbf{I}}$ denotes the unit tensor. It is seen that
Eqs.\ (\ref{3.56}) and (\ref{3.57}) coincide with Eqs.\ (\ref{1.1a})
and (\ref{1.1b}), respectively, with all vectors appearing in the
latter two equations restricted to $\mathbb{R}^{2}$, provided that
one sets
\begin{equation}
\boldsymbol{c}=\frac{m\hat{\mathbf{L}}\cdot\boldsymbol{u}}{L^{2}}
=-\frac{m\boldsymbol{u}\cdot\hat{\mathbf{L}}}{L^{2}}.
\label{3.58}
\end{equation}
Moreover, the vector representation
\begin{equation}
\boldsymbol{V}=\boldsymbol{v}+\boldsymbol{u}
\label{3.59}
\end{equation}
of Eq.\ (\ref{3.51}) is seen to match our Eq.\ (\ref{3.47}), since in
$\mathbb{R}^{2}$, by virtue of Eqs.\ (\ref{3.58}) and (\ref{3.57}),
it holds that
\begin{equation}
-\frac{\hat{\mathbf{L}}\cdot\boldsymbol{c}}{m}=\boldsymbol{u}.
\label{3.60}
\end{equation}
It is also worth noting that the authors allude to the possibility of
mapping a Keplerian ellipse onto two branches of a hyperbola, which
corresponds to the situation illustrated in Fig.\/~\ref{fig3} of the
present work; however, they do not provide any mathematical details.

We would be remiss not to point out that the analysis presented by
Kholshevnikov and Titov is not free from errors. The second pair of
equations on page 81, which in our notation is
\begin{equation}
\frac{X}{R}=\frac{x}{r},
\qquad
\frac{Y}{R}=\frac{y}{r}
\qquad (\textrm{incorrect!}),
\label{3.61}
\end{equation}
is not generally valid in light of our Eqs.\ (\ref{3.7}),
(\ref{3.35}), and (\ref{3.36}). In turn, the subsequent equation
\begin{equation}
R=r\frac{xv_{y}-yv_{x}}{xv_{y}-yv_{x}+xu_{y}-yu_{x}}
\qquad (\textrm{incorrect!}),
\label{3.62}
\end{equation}
which corresponds to our Eq.\ (\ref{3.34}), suffers from the
unjustified omission of the absolute value symbols around the
fraction on the right-hand side. Since both Eqs.\ (\ref{3.61}) and
(\ref{3.62}) are essential for deriving Eqs.\ (\ref{3.53}) and
(\ref{3.54}), the fact that the latter two turn out, as we have shown
above, to be ultimately correct should be regarded as somewhat
fortunate. Another drawback is that their discussion lacks an
examination, analogous to the one in Sec.\/~\ref{II} of this paper,
of how the combination of the derived transformations (\ref{3.53})
and (\ref{3.54}) affects Newton's equation of motion for the
Kepler--Coulomb problem.

In conclusion, we consider Kholshevnikov and Titov's approach
promising but deserving of further refinement.
%
%
\section{Conservation of the Laplace--Runge--Lenz vector as a
consequence of the dynamical quasi-symmetry}
\label{IV}
\setcounter{equation}{0}
We now demonstrate that the conservation of the Laplace--Runge--Lenz
vector in the Kepler--Coulomb problem is a direct consequence of the
dynamical quasi-symmetry revealed in Sec.\/~\ref{II}. For greater
generality, we abandon the coordinate-based notation employed in
Sec.\/~\ref{III} and return to the coordinate-free approach used in
Sec.\/~\ref{II}. We assume that the plane of motion is arbitrarily
oriented in $\mathbb{R}^{N}$ and use $\boldsymbol{r}$ to denote both
the position vector of a particle moving along its planar orbit and
the position vector of a geometric point in $\mathbb{R}^{N}$, both
defined with respect to the center of force; the intended meaning
will be clear from the context. The same convention is adopted for
the transformed position vector $\boldsymbol{R}$.

Consider a particle moving along a conic orbit of eccentricity $e$
and semi-latus rectum $p$ under the action of the force (\ref{2.1}).
Assume that the vector $\boldsymbol{c}$ is such that the line of
intersection of the plane of motion with the hyperplane
$1+\boldsymbol{c}\cdot\boldsymbol{r}=0$ has no points in common with
the particle's orbit and, if $e>1$, also with the partner hyperbolic
orbit [cf.\ the left-hand panels in Figs.\ \ref{fig1}, \ref{fig4},
and \ref{fig8}; it follows from the discussion in Sec.\ \ref{III.2.1}
that the set of all vectors $\boldsymbol{c}$ satisfying this
constraint defines an open hypercylinder in $\mathbb{R}^{N}$, which
we denote as $\mathscr{C}_{ep}$; one has
$\mathscr{C}_{ep}=D_{ep}\times\mathbb{R}^{N-2}$, where $D_{ep}$ is
the open disk characterized in local planar Cartesian coordinates by
Eq.\ (\ref{3.27})]. Then, if $\alpha<0$, the motion takes place in
the half-plane where \mbox{$1+\boldsymbol{c}\cdot\boldsymbol{r}>0$},
which contains the force center at $\boldsymbol{r}=\boldsymbol{0}$;
conversely, if $\alpha>0$, the motion remains in the half-plane
where $1+\boldsymbol{c}\cdot\boldsymbol{r}<0$. As shown in
Sec.\/~\ref{III.2}, the transformation (\ref{1.1a}) maps the
particle's orbit onto the entire ellipse when $0\leqslant
e\leqslant1$, and onto an elliptic arc when $e>1$, with the endpoints
of the arc (which are images of points at infinity on the original
hyperbolic orbit) lying on the line that is the intersection of the
plane of transformed motion with the hyperplane
\mbox{$1-\boldsymbol{c}\cdot\boldsymbol{R}=0$}. By virtue of Eq.\
(\ref{2.7}), we infer that for $\alpha<0$ the transformed trajectory
is confined to the half-plane where
\mbox{$1-\boldsymbol{c}\cdot\boldsymbol{R}\geqslant0$}, which
contains the center of the transformed-force at
\mbox{$\boldsymbol{R}=\boldsymbol{0}$}, whereas for $\alpha>0$ it
lies entirely in the half-plane where
\mbox{$1-\boldsymbol{c}\cdot\boldsymbol{R}\leqslant0$} (cf.\ the
right-hand panels in Figs.\ \ref{fig1}, \ref{fig4}, and \ref{fig8}).
Thus, in the case under consideration, the relation
\begin{equation}
\alpha\sgn(1-\boldsymbol{c}\cdot\boldsymbol{R})=-|\alpha|
\label{4.1}
\end{equation}
holds on the transformed trajectory and the resulting equation of
motion (\ref{2.17}) takes the form
\begin{equation}
m\frac{\mathrm{d}\boldsymbol{V}}{\mathrm{d}T}
=-|\alpha|\frac{\boldsymbol{R}}{R^{3}}.
\label{4.2}
\end{equation}
Taking the dot product of Eq.\ (\ref{4.2}) with $\boldsymbol{V}$, we
deduce the conservation law
\begin{equation}
\frac{\mathrm{d}\mathscr{E}}{\mathrm{d}T}=0
\label{4.3}
\end{equation}
for the quantity
\begin{equation}
\mathscr{E}=\frac{1}{2}m\boldsymbol{V}^{2}-\frac{|\alpha|}{R},
\label{4.4}
\end{equation}
which plays the role of energy in the transformed motion; in what
follows, we shall refer to $\mathscr{E}$ as pseudo-energy. Since it
holds that
\begin{equation}
\frac{\mathrm{d}\mathscr{E}}{\mathrm{d}t}
=\frac{\mathrm{d}T}{\mathrm{d}t}
\frac{\mathrm{d}\mathscr{E}}{\mathrm{d}T}=0,
\label{4.5}
\end{equation}
the pseudo-energy $\mathscr{E}$ is conserved not only with respect to
the pseudo-time $T$, but also with respect to the physical time $t$.

We now express the pseudo-energy $\mathscr{E}$ in terms of the
original dynamical variables $\boldsymbol{r}$ and $\boldsymbol{v}$.
Using successively Eqs.\ (\ref{3.47}), (\ref{1.1a}), (\ref{2.7}), and
(\ref{4.1}), we obtain
\begin{equation}
\mathscr{E}=\frac{1}{2}m\boldsymbol{v}^{2}
-\boldsymbol{v}\cdot\hat{\mathbf{L}}\cdot\boldsymbol{c}
+\frac{(\hat{\mathbf{L}}\cdot\boldsymbol{c})^{2}}{2m}
+\frac{\alpha}{r}(1+\boldsymbol{c}\cdot\boldsymbol{r}).
\label{4.6}
\end{equation}
Since the tensor $\hat{\mathbf{L}}$ is antisymmetric, it is immediate
that
\begin{equation}
\boldsymbol{v}\cdot\hat{\mathbf{L}}\cdot\boldsymbol{c}
=-\boldsymbol{c}\cdot\hat{\mathbf{L}}\cdot\boldsymbol{v}
\label{4.7}
\end{equation}
and also
\begin{equation}
(\hat{\mathbf{L}}\cdot\boldsymbol{c})^{2}
=-\boldsymbol{c}\cdot\hat{\mathbf{L}}
\cdot\hat{\mathbf{L}}\cdot\boldsymbol{c}.
\label{4.8}
\end{equation}
Once these two relations are inserted into the right-hand side of
Eq.\ (\ref{2.5}), after straightforward rearrangements we obtain
\begin{equation}
\mathscr{E}=\left(\frac{1}{2}m\boldsymbol{v}^{2}
+\frac{\alpha}{r}\right)
+\boldsymbol{c}\cdot\left(\hat{\mathbf{L}}\cdot\boldsymbol{v}
+\alpha\frac{\boldsymbol{r}}{r}\right)
-\boldsymbol{c}\cdot\frac{\hat{\mathbf{L}}
\cdot\hat{\mathbf{L}}}{2m}\cdot\boldsymbol{c}.
\label{4.9}
\end{equation}
According to Eq.\ (\ref{4.5}), the pseudo-energy $\mathcal{E}$ is a
constant of motion, with respect to $t$, for \emph{an arbitrary\/}
vector $\boldsymbol{c}\in\mathscr{C}_{ep}$. This implies that each of
the three terms on the right-hand side of Eq.\ (\ref{4.9}) must
separately be a constant of motion. The constancy of the zeroth-order
term simply expresses the conservation of energy
\begin{equation}
E=\frac{1}{2}m\boldsymbol{v}^{2}+\frac{\alpha}{r}=\textrm{const}
\label{4.10}
\end{equation}
in the original motion, which, of course, could have been deduced
from Eq.\ (\ref{2.2}) far more easily. The constancy of the quadratic
term implies that the symmetric tensor
\mbox{$\hat{\mathbf{L}}\cdot\hat{\mathbf{L}}$} is a constant of
motion, but this already follows from the conservation law
(\ref{2.4}), so again no additional information is obtained. By
contrast, the constancy of the first-order term implies a new (in the
context of this work) conservation law
\begin{equation}
\boldsymbol{A}=\hat{\mathbf{L}}\cdot\boldsymbol{v}
+\alpha\frac{\boldsymbol{r}}{r}=\textbf{const}.
\label{4.11}
\end{equation}
The vector $\boldsymbol{A}$ that appears in Eq.\ (\ref{4.11}) is an
$N$-dimensional generalization of the three-dimensional
Laplace--Runge--Lenz vector. For in three dimensions, after switching
to the cross-product notation, we have
\begin{equation}
\hat{\mathbf{L}}\stackrel{N=3}{\longrightarrow}\boldsymbol{L}
=m\boldsymbol{r}\times\boldsymbol{v}
\label{4.12}
\end{equation}
and
\begin{equation}
\hat{\mathbf{L}}\cdot\boldsymbol{v}
\stackrel{N=3}{\longrightarrow}-\boldsymbol{L}\times\boldsymbol{v}
=\boldsymbol{v}\times\boldsymbol{L},
\label{4.13}
\end{equation}
so that
\begin{equation}
\boldsymbol{A}=\boldsymbol{v}
\times\boldsymbol{L}+\alpha\frac{\boldsymbol{r}}{r}
\qquad (\textrm{in 3D}),
\label{4.14}
\end{equation}
in agreement with what can be found in standard reference texts [cf.,
e.g., Goldstein et al.\/ (2002, Sec.\/~3.9) or Kotkin et al.\/ (2024,
Sec.\/~3.3)].

Since the vector (\ref{4.11}), associated with the use of the
variables $(t,\boldsymbol{r},\boldsymbol{v})$, is conserved during
the original motion, the vector
\begin{equation}
\boldsymbol{\mathscr{A}}=\hat{\mathbf{L}}\cdot\boldsymbol{V}
-|\alpha|\frac{\boldsymbol{R}}{R},
\label{4.15}
\end{equation}
pertinent to the use of the variables
($T,\boldsymbol{R},\boldsymbol{V})$, must also be a constant of
motion. However, it appears that no conservation law in the variables
$(t,\boldsymbol{r},\boldsymbol{v})$ follows from this, other than
those already deduced above. Indeed, if the right-hand side of Eq.\
(\ref{4.15}) is transformed with the help of Eqs.\ (\ref{3.47}),
(\ref{1.1a}), (\ref{2.7}), and (\ref{4.1}), this yields
\begin{equation}
\boldsymbol{\mathscr{A}}=\left(\hat{\mathbf{L}}\cdot\boldsymbol{v}
+\alpha\frac{\boldsymbol{r}}{r}\right)
-\frac{\hat{\mathbf{L}}\cdot\hat{\mathbf{L}}}{m}\cdot\boldsymbol{c}.
\label{4.16}
\end{equation}
From this, it is seen that the time-independence of the vector
$\boldsymbol{\mathscr{A}}$ simply implies that both the vector
$\boldsymbol{A}$ and the tensor
$\hat{\mathbf{L}}\cdot\hat{\mathbf{L}}$ are conserved quantities, as
has already been shown. Examining the vector
$\boldsymbol{\mathscr{A}}$ nevertheless is not without merit, since
Eqs.\ (\ref{4.16}) and (\ref{4.9}) imply the differential relationship
\begin{equation}
\boldsymbol{\mathscr{A}}
=\frac{\partial\mathcal{E}}{\partial\boldsymbol{c}},
\label{4.17}
\end{equation}
which is of intrinsic interest.
%
%
\section{Final remarks}
\label{V}
\setcounter{equation}{0}
We close this work with several additional remarks.

First, the analysis carried out in Sec.\/~\ref{III} has been
restricted to the case of non-zero angular momentum, $L\neq0$, when
both the original orbits and transformed curves are non-degenerate
conics (or their subsets). The omitted special case $L=0$ corresponds
to both the original and transformed motions taking place along conics
degenerating into line segments or half-lines; its independent
analysis should not pose any difficulty for the interested reader.

Second, at the beginning of Sec.\ \ref{3.1}, we have taken the
standardized polar representation (\ref{3.1}) of non-degenerate
Kepler--Coulomb conic orbits as given. The rationale for doing so has
been that for $N=3$ Eq.\ (\ref{3.1}) is a well-established textbook
knowledge, while for $N$ arbitrary it is one of results from the
paper by {\"O}nder and Ver{\c{c}}in (2006). However, it is worth
observing [cf.\ the remark on page 5740 of Krause (1994)] that even
Eq.\ (\ref{3.1}) itself can be derived by applying a special case of
the transformation (\ref{1.1a}) to a particular solution of Newton's
equation (\ref{2.2}) describing uniform circular motion under the
action of an attractive Kepler--Coulomb force.

Third, the group defined by Eqs.\ (\ref{1.1a}) and (\ref{1.1b}) is a
subgroup of the group expressed by
\begin{subequations}
\begin{equation}
\boldsymbol{r}\to\boldsymbol{R}=\frac{\lambda\mathsf{O}\boldsymbol{r}}
{1+\boldsymbol{c}\cdot\boldsymbol{r}}
\label{5.1a}
\end{equation}
and
\begin{equation}
\mathrm{d}t\to\mathrm{d}T=\frac{\lambda^{3/2}\mathrm{d}t}
{(1+\boldsymbol{c}\cdot\boldsymbol{r})^{2}},
\label{5.1b}
\end{equation}
\label{5.1}%
\end{subequations}
where $\lambda\in\mathbb{R}_{+}$, $\mathsf{O}\in\mathrm{O}(N)$,
$\boldsymbol{c}\in\mathbb{R}^{N}$. This larger group is jointly
generated by spatial rotations
\begin{equation}
\boldsymbol{r}\to\boldsymbol{R}=\mathsf{O}\boldsymbol{r},
\qquad
t\to T=t,
\label{5.2}
\end{equation}
spatio-temporal scalings
\begin{equation}
\boldsymbol{r}\to\boldsymbol{R}=\lambda\boldsymbol{r},
\qquad
t\to T=\lambda^{3/2}t,
\label{5.3}
\end{equation}
and precisely the transformations (\ref{1.1a}) and (\ref{1.1b}).
Since the transformations specified in Eqs.\ (\ref{5.2}) and
(\ref{5.3}) are symmetries of Newton's equation (\ref{2.2}), whereas 
the transformations specified in Eqs.\ (\ref{1.1}) are only 
quasi-symmetries, the group defined by Eqs.\ (\ref{5.1}) is itself a 
dynamical quasi-symmetry group of the classical Kepler--Coulomb 
problem [not a dynamical symmetry group, as claimed by Krause 
(1994)], in the sense given at the end of Sec.\ \ref{II}. Krause 
(1994) and Bor and Jackman (2023) rightly pointed out that the 
spatial parts of the transformations belonging to the group of 
spatial transformations (\ref{5.1a}) constitute a symmetry group of a 
family of conics with a common focus at 
$\boldsymbol{r}=\boldsymbol{0}$. However, it would be erroneous to
conclude from the foregoing that this is a symmetry group of the
Kepler--Coulomb \emph{orbits}. For, as we know from Sec.\ 
\ref{III.2}, the transformations from the subgroup (\ref{1.1a}) of 
the group (\ref{5.1a}) can, under certain conditions, map a 
Kepler--Coulomb orbit onto a union of disjoint conic arcs; because of 
this disjointness, such a union is \emph{not\/} itself an orbit. In 
this context, the reader may wish to consult, e.g., Fig.\/~3, where 
an elliptic orbit is mapped onto the entire hyperbola, which is a 
union of two hyperbolic orbits, one attractive and another repulsive, 
or Fig.\/~12, where an attractive hyperbolic orbit is mapped onto the 
union of a repulsive hyperbolic orbit and a semi-infinite arc of an 
attractive hyperbolic orbit.

Finally, it should be emphasized that the dynamical quasi-symmetry
discussed in this work is completely independent of the SO$(N+1)$ and
SO$(N,1)$ symmetries mentioned in the Introduction, which are
characteristic of particle states with negative and positive
energies, respectively. Transformations belonging to these two
symmetry groups preserve the particle's total energy while altering
its angular momentum tensor. Conversely, transformations from the
quasi-symmetry group alter the particle's energy, leaving its angular
momentum tensor unchanged. The SO$(N+1)$ and SO$(N,1)$ symmetries, on
the one hand, and the quasi-symmetry in question, on the other hand,
complement each other.
%
%
\section*{Acknowledgment}
Access to Mathematica software, granted by CI TASK, is acknowledged.
\section*{Data availability}
No data were created or analyzed in this study. All figures can be
reproduced directly from the presented equations.
\section*{Conflict of Interest Statement}
The author has no relevant financial or non-financial interests to
disclose.
%
%
\section*{References}
\begin{list}{}{}
\item \hspace*{-2.5em}
   Appell, P.:
   De l'homographie en m{\'e}canique,
   Compt.\ Rend.\ Acad.\ Sci.\ Paris 108, 224--226 (1889a).
   \newline
   [\href{https://gallica.bnf.fr/ark:/12148/bpt6k3064g/f224.item}%
   {https:/\!/gallica.bnf.fr/ark:/12148/bpt6k3064g/f224.item}]
\item \hspace*{-2.5em}
   Appell, P.:
   De l'homographie en m{\'e}canique,
   Am.\ J.\ Math.\ 12, 103--114 (1889b).
   \newline
   [\href{https://doi.org/10.2307/2369252}%
   {https:/\!/doi.org/10.2307/2369252}]
\item \hspace*{-2.5em}
   Bor, G., Jackman, C.:
   Revisiting Kepler: new symmetries of an old problem,
   Arnold Math.\ J.\ 9, 267--293 (2023).
   \newline
   [\href{https://doi.org/10.1007/s40598-022-00213-2}%
   {https:/\!/doi.org/10.1007/s40598-022-00213-2}]
\item \hspace*{-2.5em}
   Cari{\~n}ena, J.\ F., L{\'o}pez, C., del Olmo, M.\ A., Santander,
   M.:
   Conformal geometry of the Kepler orbit space,
   Celest.\ Mech.\ Dyn.\ Astron.\ 52, 307--343 (1991).
   \newline
   [\href{https://doi.org/10.1007/BF00048449}%
   {https:/\!/doi.org/10.1007/BF00048449}]
\item \hspace*{-2.5em}
   Cordani, B.:
   The Kepler Problem:\ Group Theoretical Aspects, Regularization and
   Quantization, with Application to the Study of Perturbations,
   Birkh{\"a}user, Basel (2003).
   \newline
   [\href{https://doi.org/10.1007/978-3-0348-8051-0}%
   {https:/\!/doi.org/10.1007/978-3-0348-8051-0}]
\item \hspace*{-2.5em}
   Goldstein, H., Poole, C., Safko, J.:
   Classical Mechanics, 3rd ed.,
   Addison-Wesley, San Francisco (2002).
\item \hspace*{-2.5em}
   Gy{\"o}rgyi, G.:
   Kepler's equation, Fock variables, Bacry's generators and Dirac
   brackets,
   Nuovo Cimento A 53, 717--736 (1968).
   \newline
   [\href{https://doi.org/10.1007/BF02721718}%
   {https:/\!/doi.org/10.1007/BF02721718}]
\item \hspace*{-2.5em}
   Gy{\"o}rgyi, G.:
   Kepler's equation, Fock variables, Bacry's generators and Dirac
   brackets. II. --- Classical and quantum group dynamics of the
   Kepler problem,
   Nuovo Cimento A 62, 449--474 (1969).
   \newline
   [\href{https://doi.org/10.1007/BF02753005}%
   {https:/\!/doi.org/10.1007/BF02753005}]
\item \hspace*{-2.5em}
   Kholshevnikov, K.\ V., Titov, V.\ B.:
   The Two-Body Problem,
   Saint Petersburg State University, Saint Petersburg (2007),
   Sec.\ 2.5 (in Russian).
\item \hspace*{-2.5em}
   Kotkin, G.\ L., Serbo, V.\ G., Chernykh, A.\ I.:
   Lectures on Analytical Mechanics,
   Oxford University Press, Oxford (2024).
   \newline
   [\href{https://doi.org/10.1093/oso/9780198894674.001.0001}%
   {https:/\!/doi.org/10.1093/oso/9780198894674.001.0001}]
\item \hspace*{-2.5em}
   Mariwalla, K.:
   Dynamical symmetries in mechanics,
   Phys.\ Rep.\ 20, 287--362 (1975).
   \newline
   [\href{https://doi.org/10.1016/S0370-1573(75)80001-9}%
   {https:/\!/doi.org/10.1016/S0370-1573(75)80001-9}]
\item \hspace*{-2.5em}
   Milnor, J.:
   On the geometry of the Kepler problem,
   Am.\ Math.\ Month.\ 90, 353--365 (1983).
   \newline
   [\href{https://doi.org/10.1080/00029890.1983.11971232}%
   {https:/\!/doi.org/10.1080/00029890.1983.11971232}]
\item \hspace*{-2.5em}
   {\"O}nder, M., Ver{\c{c}}in, A.:
   Orbits of the $n$-dimensional Kepler--Coulomb problem and
   universality of the Kepler laws,
   Eur.\ J.\ Phys.\ 27, 49--55 (2006).
   \newline
   [\href{https://doi.org/10.1088/0143-0807/27/1/005}%
   {https:/\!/doi.org/10.1088/0143-0807/27/1/005}]
\item \hspace*{-2.5em}
   Osipov, Yu.\ S.:
   The Kepler problem and geodesic flows in spaces of constant
   curvature,
   Celest.\ Mech.\ 16, 191--208 (1977).
   \newline
   [\href{https://doi.org/10.1007/BF01228600}%
   {https:/\!/doi.org/10.1007/BF01228600}]
\item \hspace*{-2.5em}
   Szmytkowski, R.:
   Comment on: ``On the complete symmetry group of the classical
   Kepler system'' [J.\ Math.\ Phys.\ 35, 5734 (1994)],
   to be submitted.
\item \hspace*{-2.5em}
   Wheeler, J.\ T.:
   Not-so-classical mechanics:\ unexpected symmetries of classical
   motion,
   Can.\ J.\ Phys.\ 83, 91--138 (2005).
   \newline
   [\href{https://doi.org/10.1139/p05-003}%
   {https:/\!/doi.org/10.1139/p05-003}]
\item \hspace*{-2.5em}
   Wulfman, C.\ E.:
   Dynamical Symmetry,
   World Scientific, Hackensack, NJ (2011).
   \newline
   [\href{https://doi.org/10.1142/7548}%
   {https:/\!/doi.org/10.1142/7548}]
\end{list}
\end{document}